\documentclass[usenatbib]{mnras}

\usepackage[english]{babel}
\usepackage[utf8x]{inputenc}
\usepackage[T1]{fontenc}

\usepackage{amssymb}
\usepackage{amsmath}
\usepackage{cuted}
\usepackage{bm}
\usepackage{booktabs}
\usepackage{gensymb}
\usepackage{graphicx}
\usepackage{hyperref}
\usepackage{mathtools}
\usepackage{multirow}
\usepackage{orcidlink}
\usepackage{textcomp}
\usepackage[para,online,flushleft]{threeparttable}
\usepackage{ulem}
\usepackage{url}
\usepackage[table,xcdraw]{xcolor}

\newcommand{\sneia}{SNe~Ia}
\newcommand{\hsf}{Hawai`i Supernova Flows}
\newcommand{\rsim}{r_{\text{sim}}}
\newcommand{\Om}{\Omega_\text{M}}
\newcommand{\zcos}{z_\text{cos}}
\newcommand{\zobs}{z_\text{obs}}

\title[HSF: Bulk Flows]{Hawai`i Supernova Flows: Bulk Flow Measurements using SNe Ia in the Optical and NIR }
\author[Do et al.]{Aaron Do$^1$\thanks{E-mail:\href{mailto:ajmd6@cam.ac.uk}{ajmd6@cam.ac.uk}}\orcidlink{0000-0003-3429-7845},
Kaisey S. Mandel$^{1,2}$\orcidlink{0000-0001-9846-4417},
Benjamin J. Shappee$^3$\orcidlink{0000-0003-4631-1149},
R. Brent Tully$^3$\orcidlink{0000-0002-9291-1981}, 
John L. Tonry$^3$\orcidlink{0000-0003-2858-9657}, 
\newauthor{
David Rubin$^{3,4}$\orcidlink{0000-0001-5402-4647},
David O. Jones$^5$\orcidlink{0000-0002-6230-0151}, 
Mitchell Dixon$^5$\orcidlink{0000-0003-0928-0494}, 
Thomas de Jaeger$^6$\orcidlink{0000-0001-6069-1139}, 
Dan Scolnic$^7$\orcidlink{0000-0002-4934-5849},
}
\newauthor{
Erik R. Peterson$^{8,9}$\orcidlink{0000-0001-8596-4746}, 
Christopher R. Burns$^{10}$\orcidlink{0000-0003-4625-6629} 
}\\
$^1$Institute of Astronomy and Kavli Institute for Cosmology, University of Cambridge, Madingley Road, Cambridge CB3 0HA, UK\\
$^{2}$Statistical Laboratory, DPMMS, University of Cambridge, Wilberforce Road, Cambridge CB3 0WB, UK\\
$^3$Institute for Astronomy, University of Hawai`i, 2680 Woodlawn Dr., Honolulu, HI 96822, USA\\
$^4$Department of Physics and Astronomy, University of Hawai‘i, Honolulu, HI 96822, USA\\
$^5$Institute for Astronomy, University of Hawai‘i, 640 N Aohoku Place, Hilo, HI 96720, USA\\
$^6$CNRS/IN2P3, Sorbonne Universit\'{e}, Universit\'{e} Paris Cit\'{e}, Laboratoire de Physique Nucl\'{e}aire et de Hautes \'{E}nergies, 75005,\\ Paris, France\\
$^7$Department of Physics, Duke University, Durham, NC 27708, USA\\
$^8$Department of Physics, University of Michigan, Ann Arbor, MI 48109, USA\\
$^9$Society of Fellows, University of Michigan, Ann Arbor, MI 48109, USA\\
$^{10}$The Observatories of the Carnegie Institution for Science, 813 Santa Barbara Street, Pasadena, CA 91101, USA\\
}
\date{Accepted XXX. Received YYY; in original form ZZZ}
\pubyear{\the\year{}}

\begin{document}
\maketitle{}
\label{firstpage}
\pagerange{\pageref{firstpage}--\pageref{lastpage}}

\begin{abstract}
The present day peculiar velocity-field was sourced by primordial density fluctuations and sculpted over the lifespan of the Universe.
Cosmological models such as $\Lambda$CDM make predictions for various statistical properties of peculiar velocities.
Bulk flow, the average velocity within a given volume, has an expectation value of $\vec{0}$ due to isotropy, and a variance directly tied to the Hubble constant, the growth-rate of structure, and the matter power spectrum.
In this paper, we use the redshifts and optical and near-infrared distance estimates to Type Ia Supernovae (SNe~Ia) within subsets of the Hawai`i Supernova Flows dataset to infer the bulk flow within $z \lesssim 0.1$.
The inferred speeds vary between $\sim100$ to $400$ km s$^{-1}$ but are all consistent with the predictions of $\Lambda$CDM.
As a secondary focus, we discuss the systematic uncertainty introduced by the discrete choice of methodology using two bulk flow estimators, two types of SN~Ia distance estimators, and data covering two distinct regimes in wavelength space.
\end{abstract}

\section{Introduction}
\label{sec:introduction}

The cosmological principle posits that the Universe is homogeneous and isotropic at large enough scales.
By definition, the integrated directional components of an isotropic vector field must cancel out such that the average vector has a magnitude of 0.
Thus, the sample average of a velocity field within a certain volume, called bulk flow, should tend towards the population mean of 0 as the volume expands and the sample size increases.
In a cosmological context, bulk flow describes the average of the peculiar velocity field, which is induced by gravitational over- and under-densities.
The measured ``velocities'' due to the expansion of the Universe cancel out when sampled isotropically, otherwise providing a monopole-like term to the measured bulk flow that can be corrected for based on the sampled volume.
The current model that best matches our observations of the Universe, dark energy and cold dark matter ($\Lambda$CDM), describes how the statistical properties of bulk flow should evolve as a function of the length-scale of the enclosed volume.
Measurements of bulk flows generally agree with the predictions of $\Lambda$CDM at relatively small length-scales \citep[$R \lesssim 70 ~h^{-1}\text{ Mpc}$;][]{ma14, hoffman15, scrimgeour16, qin18, qin21}, but have been observed in excess of the predicted values at larger length-scales \citep[$70~h^{-1}\text{ Mpc} < R < 300~h^{-1}\text{ Mpc}$;][]{kashlinsky08, watkins15, peery18, howlett22, watkins23, whitford23}.
This is not a clean division, as some studies find disagreement with $\Lambda$CDM at small length-scales \citep[e.g.][]{watkins09, appleby15}, or consistency at large ones \citep[e.g.,][]{nusser11, turnbull12, feindt13}.
The precision of these discrepant measurements is still low enough for other mismatches between theory and observation to present more serious challenges to the current paradigm \citep{peebles22}.
However, as extragalactic surveys become more ambitious and measurements become more precise, excess bulk flow measurements at large scales may grow in significance.

\citet{nusser16} lists two subtly different definitions that are often conflated in studies of bulk flow.
Given a peculiar velocity field $\vec{v}(\vec{r})$ as a function of comoving distance vector $\vec{r}$, one can define the Cartesian unit vector $\hat{x}^i$ ($i =$ 1, 2, and 3) projections of the bulk flow $\vec{B}$ evaluated within volume $V$ as
\begin{equation}
    \label{eqn:bulk}
    \vec{B}_\text{I}\cdot\hat{x}^i \equiv \int_V \mathrm{d}^3r\, g(\vec{r})\vec{v}(\vec{r})\cdot\hat{x}^i
\end{equation}
where $g(\vec{r})$ is a weighting function normalizing the volume integral as $\int_V \mathrm{d}^3r g(\vec{r}) = 1$.
Alternatively, one could consider only the radial projection of the peculiar velocity field, $u(\vec{r}) \equiv \vec{v}(\vec{r})\cdot\hat{r}$ and use the definition
\begin{equation}
    \label{eqn:bulk_alt}
    \vec{B}_\text{II}\cdot\hat{x}^i \equiv 3\int_V \mathrm{d}^3r\,g(\vec{r})u(\vec{r})\hat{r}\cdot\hat{x}^i.
\end{equation}
The two definitions are equivalent when
\begin{equation}
    \vec{v}(\vec{r})\cdot\hat{x}^i = 3(\vec{v}(\vec{r})\cdot\hat{r}) (\hat{r}\cdot \hat{x}^i)
\end{equation}
which is not guaranteed, but occurs if the peculiar velocity field is constant throughout the volume ($\vec{v}(\vec{r}) = \vec{B}_0$), or if specific weight functions are chosen under certain conditions \citep{nusser14}.
Given that the definition in Equation \ref{eqn:bulk_alt} is more aligned with observable quantities, all future references to bulk flows use that definition ($\vec{B} \equiv \vec{B_\text{II}}$).

In this paper, we present a measurement of bulk flow using Type Ia supernovae (\sneia{}) observed in the optical and near-infrared (NIR) regimes as part of the \hsf{} project \citep{do25}.
The paper is organized as follows.
In Section \ref{sec:bulk_flow_review} we review bulk flows, covering both the predictions from concordance cosmology and the methods developed to overcome the many systematic challenges in measuring bulk flows.
In Section \ref{sec:HSF_data}, we briefly describe the \hsf{} data, cuts, and samples, before evaluating the performance of several bulk flow estimation methods against comparable simulations in Section \ref{sec:validation}.
We present our measurement of bulk flow using the \hsf{} data in Section \ref{sec:results}.
We discuss those results in the context of other studies and future datasets in Section \ref{sec:discussion}.

\subsection*{Notation}
The $i$ and $j$ subscripts and superscripts are reserved for indexing the three Cartesian axes.
The $n$ and $m$ subscripts are reserved for indexing individual tracers within a total sample of $N$ tracers.
Unitalicised subscripts are descriptive and are placed before variable subscripts (e.g., $z_{\text{obs},n}$ is the observed redshift of the $n$th tracer).

Variables representing scalar quantities have no formatting beyond the italicisation common to all variables.
Vector quantities are denoted with an arrow.
If the same variable used for a vector appears without an arrow, it represents the vector's scalar magnitude.
Unit vectors are represented with hats.
Matrices are bolded and may be implied by indexed matrix elements.
As an example, the rank-$N$ identity matrix is implied by $\bm{I} = \delta^K_{mn}$ where $\delta^K_{mn}$ is the Kronecker delta function.

Normal distributions are parametrized by a mean and variance, e.g., $x \sim \mathcal{N}(\mu,\sigma^2)$.
\section{Reviewing Bulk Flows}
\label{sec:bulk_flow_review}
\subsection{Predictions from Theory}
\label{sec:lcdm}

In the scope of this paper, we assume the cosmological principle holds (there exists a transitional length scale above which the Universe is homogeneous and isotropic), the Universe is described by a Friedmann–Lemaître–Robertson–Walker (FLRW) metric, and the spatial curvature density $\Omega_K$ is 0.
Under these assumptions, the peculiar velocity field can be decomposed into linear and non-linear components, with the former exclusively sourced by gravitational instabilities growing from Gaussian density fluctuations via
\begin{equation}
    \vec{v}(\vec{r}) = -\nabla\Phi(\vec{r})
    \label{eqn:v_density}
\end{equation}
where $\Phi$ is a scalar potential field \citep{bertschinger89}.
The non-linear component comprises all other sources of peculiar velocity (e.g., Virial motion, baryonic effects, pressure, etc.) and is confined to relatively small spatial scales ($\lesssim 10$ Mpc).
Assuming the linear component dominates above length scale $R$, the velocity field can be smoothed with a ``window'' or ``filter'' function $W(R)$.
The bulk flow is then a Gaussian random variate with a mean of 0 and a variance \citep[Equation 18.3.4;][]{coles02}
\begin{equation}
    \label{eqn:lcdm_pv_variance}
    \sigma^2(R) = \frac{H^2_0 f^2}{2\pi^2} \int_{0}^\infty P(k) \tilde{W}^2(k,R)\mathrm{d}k
\end{equation}
where $H_0$ is the Hubble constant, $f$ is the growth rate, $P(k)$ is the matter power spectrum as a function of wavenumber $k$, and $\tilde{W}(k,R)$ is the Fourier transform of the window function at $k$.
The Hubble constant, growth rate, and matter power spectrum are defined by a given cosmological model, but the window function is model-agnostic.
Typical choices for isotropic window functions include the spherical top-hat
\begin{equation}
    \tilde{W}(k,R) = \frac{3(\sin{kR} -kR\cos{kR})}{(kR)^3}
    \label{eqn:tophat_window}
\end{equation}
or Gaussian
\begin{equation}
    \tilde{W}(k,R) = \exp{\frac{-k^2R^2}{2}}.
    \label{eqn:gaussian_window}
\end{equation}

The variance expected given a catalogue of $N$ tracers can be calculated by replacing the isotropic $\tilde{W}^2(k,R)$ with a tensor angle-averaged window function \citep{feldman10, qin18, watkins23}:
\begin{equation}
    \label{eqn:data_window_function}
    \tilde{{W}}^2_{ij}(k) = \sum_{m=1}^N \sum_{n=1}^N (\vec{w}_{n} \cdot \hat{x}^i) (\vec{w}_{m} \cdot \hat{x}^j) f_{nm}(k)
\end{equation}
where the $N$ weight vectors are defined by the specific bulk flow estimation method used (see Section \ref{sec:measuring_bfs}) and $\bm{f}(k)$ (not to be confused with growth rate $f$) is the correlated window function \citep[Equation 30;][]{kaiser88}, which is formulated as an analytic function of $\vec{r}_n$ and $\vec{r}_m$ in the appendix of \citet{ma11}.
Both formulations are reproduced here for convenience:
\begin{equation}
    \begin{split}
    f_{nm}(k) =& \int \frac{\mathrm{d}^2 \hat{k}}{4\pi} (\hat{r}_n \cdot \hat{k})(\hat{r}_m \cdot \hat{k}) \times \exp\left[ik\hat{k} \cdot (\vec{r}_n - \vec{r}_m)\right] \\
    =& \frac{\cos \alpha }{3}\left[j_0(kA) - 2j_2(kA)\right] + \frac{r_n r_m \sin^2\alpha}{A^2}j_2(kA)
    \end{split}
    \label{eqn:corr_window_fn}
\end{equation}
where $\alpha$ is the relative angle between $\hat{r}_n$ and $\hat{r}_m$, $A$ is the magnitude of their difference vector $|\vec{r}_n - \vec{r}_m| = \sqrt{r^2_n + r^2_m - 2r_n r_m \cos \alpha}$, and $j_0$ and $j_2$ are 0th and 2nd order spherical Bessel functions of the first kind.
Substituting the window function $\tilde{W}^2(k,R)$ in Equation \ref{eqn:lcdm_pv_variance} with the one in Equation \ref{eqn:data_window_function} yields a covariance matrix $\bm{\Sigma}_{\text{cos}}$ rather than a scalar function of $R$.
The covariance matrix is a function of tracer distances, sky positions, and projected weights, the last term being sensitive to the former terms as well as tracer redshifts and uncertainties.

Given that the peculiar velocity distribution of dark matter halos is Maxwellian \citep{li12}, the probability distribution of measuring a bulk flow $|\vec{B}| = V$ is \citep[Equation 18.3.2;][]{coles02}:
\begin{equation}
    P(V)\,\mathrm{d}V = \sqrt{\frac{54}{\pi}}\left(\frac{V}{\sigma}\right)^{2}\exp{\left[-\frac{3}{2}\left(\frac{V}{\sigma}\right)^2\right]}\frac{\mathrm{d}V}{\sigma}.
    \label{eqn:V_prob_dist}
\end{equation}

\subsection{Inferring Peculiar Velocities}
\label{sec:measuring_pvs}
The peculiar velocity field is three-dimensional, but extragalactic distances make measurements of transverse velocities infeasible.
Thus, inferences of the field must be made using only measurements of its radial component.
The apparent deficit in dimensionality is overcome by treating the field as curl-less, which one would expect from the growing modes of linear instability theory \citep{peebles80}.
The scalar field $\Phi(\vec{r})$ from Equation \ref{eqn:v_density} can be uniquely constrained to a global additive constant by the radial component of the peculiar velocity field.
Peculiar velocity surveys are more than sufficient for measuring bulk flows \citep{nusser14}, and indeed can be used to recover the entire three-dimensional velocity field \citep[e.g.,][]{springob14, tully16, graziani19, tully23}.

The peculiar velocity of the $n$th tracer is calculated as \citep{davis14}
\begin{equation}
    \label{eqn: pv_rad}
    1 + z_{\text{pv},n}
    = 1 +\frac{u(\vec{r}_n)}{c}
    = \frac{1 + z_{\text{obs},n}}{1 + \zcos (\vec{r}_n)}
\end{equation}
where $c$ is the speed of light in a vacuum, $\vec{r}_n$ is the $n$th tracer's comoving distance vector, $z_{\text{obs},n}$ is its observed redshift with respect to the cosmic microwave background (CMB), and $\zcos (r_n)$ is the cosmological redshift expected at $\vec{r}_n$ under the assumption of isotropy.
The comoving distance-redshift relation in a flat universe with negligible radiation density is an integral with no analytic representation for non-zero dark energy densities
\begin{equation}
    r(\zcos,H_0,\Om) = \frac{c}{H_0}
    \int_0^{\zcos} \frac{\mathrm{d}z'}{\sqrt{\Om(1+z')^3 + (1-\Om)}}.
    \label{eqn:distance-redshift}
\end{equation}
We avoid numerical integration by using the Pad\'{e} approximant \citep{aviles24} implemented with the \texttt{wcosmo} package\footnote{\url{https://github.com/ColmTalbot/wcosmo}.} \citep[originally developed for][]{talbot25}.
The analytic approximant can be inverted to give $\zcos(r,~H_0,~\Om)$, but in practice, astronomical distances are usually measured as distance moduli, which are functions of luminosity distances $d_\text{L}$
\begin{equation}
    \mu \equiv m - M = 5\log_{10}\left(\frac{d_\text{L}}{1~\text{Mpc}}\right) + 25
    \label{eqn:dist_mod}
\end{equation}
where $m$ is an apparent magnitude and $M$ is the corresponding absolute magnitude.
Given Equations \ref{eqn:distance-redshift}, \ref{eqn:dist_mod}, and the relationship $d_\text{L} = r(1+\zcos)$, a measured distance modulus with Gaussian uncertainties translates to an inferred $\zcos$ with roughly log-normal uncertainties.
The radial peculiar velocity in Equation \ref{eqn: pv_rad} can be inferred by combining assumed values of $H_0$ and $\Om$ with measured values of $\zobs$ and $\mu$.

The peculiar velocity of each tracer is the sum of the motion described by linear theory and the non-linear effects that occur at small scales (e.g., strong gravitational encounters, baryonic effects).
\citet{li12} showed that halos in a high-resolution N-body simulation moved as described by linear theory, which implies that non-linear effects amount to a noise term at smaller scales that can be suppressed by smoothing the velocity field through convolution \citep{colin11, feindt13, springob16} or grouping gravitationally associated tracers \citep{hoffman21, howlett22, watkins23}.
Additionally, the noise can be modelled as an additional layer of uncertainty \citep{kaiser88, watkins09, turnbull12, nusser14, graziani19, howlett22, watkins23} that can be calibrated against simulations \citep{carrick15, hollinger21}.

We list systematic effects that must be taken into account when inferring luminosity distances in Appendix \ref{appendix:univ_dist_biases}, and apply the corrections described therein to infer the peculiar velocities analysed in this work.

\subsection{Inferring Bulk Flows}
\label{sec:measuring_bfs}
The volume-integral expression for bulk flow in Equation \ref{eqn:bulk_alt} can be approximated with a discrete weighted average
\begin{equation}
    \vec{B} \cdot \hat{x}^i \equiv 3\int_V \mathrm{d}^3rg(\vec{r})u(\vec{r})\hat{r}\cdot\hat{x}^i \approx \sum_n^N (\vec{w}_{n} \cdot \hat{x}^i) u_n
    \label{eqn:bulk_flow_sum}
\end{equation}
where $\vec{w}_n$ is the same weight term shown in Equation \ref{eqn:data_window_function}.
The simple weighting scheme $\vec{w}_n = 3\hat{r}_n / N$ causes the sum to converge towards the integral as $N$ increases if the spatial-distribution of tracers is proportional to $g(\vec{r})$ and $u_n$ is unbiased throughout the volume.
However, this is an unrealistic, special case.
With any coupling between tracers, baryonic matter, and large-scale structure (LSS), regions with greater mass density and greater peculiar velocity will be over-represented in an observed sample, necessitating a decrease in weight to avoid bias.

There are several methodologies for inferring more widely-applicable weight terms, two of which are compared in \citet{whitford23} along with their variations.
These are the Kaiser Maximum Likelihood Estimator \citep[MLE; ][]{kaiser88} and its variant \citep[Nusser MLE;][]{nusser14}, and the Minimum Variance Estimator \citep[MVE;][]{watkins09} and its variant \citep[Peery MVE;][]{peery18}.
\citet{whitford23} test these methods against the CosmicFlows-4 dataset \citep{tully23} and a number of mocks corresponding to subsets of their data \citep{qin19, qin21, howlett22}.
They found the Nusser MLE and Peery MVE methods define weighting terms that provide accurate estimates of the bulk flow moments even with spatially-variable peculiar velocity fields and that the MVE methods can accommodate arbitrary distributions of tracers.
The analytic expressions for the weights can be found in their corresponding references.
The relevant code used in the present work was modified from the code\footnote{\url{https://github.com/abbew25/Measuring_bulkflows}.} accompanying \citet{whitford23}, the original version hereafter referred to as the ``Whitford code.''
Appendix \ref{appendix:MVE_mod} describes several systematic issues we identified and the modifications we implemented to avoid them.

As an indirect approach, if one can infer the full three-dimensional velocity field, bulk flow can be trivially integrated through Equation \ref{eqn:bulk}.
Several Bayesian reconstruction methods have been tested against simulations to show that catalogues of distance moduli, redshifts, and sky positions are sufficient for reconstructing the full velocity field.
These methods include VelocIty Reconstruction using Bayesian Inference Software \citep[VIRBIUS;][]{lavaux16}, Wiener filter and constrained realizations with Bias Gaussian Corrections \citep[BGc;][]{hoffman21}, HAmiltonian Monte carlo reconstruction of the Local EnvironmenT \citep[HAMLET;][]{valade22}, and several other algorithms \citep{boruah22, prideaux-ghee23}.

Although the connection between peculiar velocity surveys and bulk flow inference is mathematically rigorous, practical limitations can produce systematic biases beyond those mentioned in Appendix \ref{appendix:univ_dist_biases} that must be taken into account.
Over sufficiently large survey volumes one would expect the average peculiar velocity to tend towards 0, but there are astrophysical and systematic reasons why any given survey may find a non-zero average.
A local overdensity (underdensity) near the centre of a survey volume will produce a global inflow (outflow), but the same measurements could be reproduced by assuming a value for $H_0$ that is too low (high).
If the survey volume is not isotropic, this peculiar velocity monopole will not cancel out and will ``leak'' into the inferred bulk flow.
This contribution is valid for that specific survey volume if astrophysically sourced, but erroneous if due to inaccurate assumptions or if used to infer a bulk flow over a symmetric survey volume.
This issue is explicitly addressed in the Peery MVE estimator, which includes a constraint designed to enforce invariance to first-order over any discrepancies between the assumed and actual values of $H_0$.
The trade-off for this robustness is reduced sensitivity to the effects of a local void or an externally-sourced tidal field aligned with the non-spherically-symmetric regions of the survey volume.

There are also statistical limitations that can bias inferences depending on the bulk flow estimator used.
\citet{andersen16} found that when a peculiar velocity dataset comprises fewer than $\sim 500$ samples the magnitudes of the bulk flows inferred with the MLE or MVE methods tended to be overestimated, with the bias being greater for non-spherical survey geometries.
Similarly, Bayesian techniques for reconstructing the full velocity-field only depart from their priors when informed by data, but peculiar velocity measurements tend to have low signal-to-noise ratios (SNRs) and the survey volume is often sparsely populated.
This produces bulk flow estimates strongly informed by prior information, making it difficult to tell if the data support the assumed cosmological model or are indicative of an alternative model but too weak to present compelling evidence.
\section{Hawai`i Supernova Flows Data}
\label{sec:HSF_data}

\hsf{} is a project designed to map the distribution of mass in the local universe by measuring the peculiar velocity field using optical and near-infrared (NIR) observations of Type Ia Supernovae (\sneia{}).
SNe Ia-based distance measurements have been shown to be more precise when including NIR observations \citep[e.g.,][]{wood-vasey08, mandel11, kattner12, barone-nugent12, stanishev18, avelino19, mandel22, thorp22, galbany23, jones22, peterson23}, but small sample sizes have limited their use in measuring bulk flows.
The first data release of the \hsf{} project contains over four hundred distances to spectroscopically classified \sneia{} and spectroscopic redshifts for their host-galaxies.
The data are divided into samples based on both distance-measurement methodology and the wavelength range of the observations, enabling an initial study of how NIR data affects bulk flow measurements.
Detailed descriptions of the techniques used to produce the \hsf{} samples may be found in \citet{do25}.
In this section, we provide a brief review of the \hsf{} data,\footnote{Available at \url{https://github.com/ado8/HSF_DR1}.} focusing on sample definition and the treatment of various biases.

\subsection{Sub-sample Definition}
This study uses two of the three primary samples from \hsf{}: the first based on EBV\_model2 from the SuperNovae in Object Oriented Python package \citep[SNooPy; ][]{burns11, burns18} and the second based on the Spectral Adaptive Light curve Template third-generation NIR-capable model \citep[SALT3-NIR; ][]{kenworthy21, pierel22}.
The samples are respectively called SNPY\_EBV and SALT, and comprise 357 and 362 \sneia{}.
There are 399 unique \sneia{} between the two samples, with 320 appearing in both samples.
The sample is defined by survey-wide cuts, model-specific cuts, and outlier detection algorithms, all of which are tabulated in \citet{do25} and reproduced here.
The survey-wide cuts retain only targets that are spectroscopically classified as \sneia{} (cutting several peculiar subtypes), associated with galaxies with spectroscopic redshifts, have Galactic extinction $E(B-V)_\text{MW} < 0.3$ mag according to the map of \citet{schlafly11}, were observed at least 5 times, and have successful scene-modelled photometry.
The model-specific cuts for the SNPY\_EBV sample retain \sneia{} with colour-stretch parameters $0.6 < s_{BV} < 1.3$ and colour-stretch uncertainties $\sigma_{s_{BV}} < 0.2$, matching the ``loose'' cuts from \citet{jones22}.
Only \sneia{} with host-galaxy extinction values $E(B-V)_\text{host} < 0.3$ mag are retained in the SNPY\_EBV sample.
The model-specific cuts for the SALT sample retain \sneia{} with shape parameter $|x_1| < 3$, shape uncertainty $\sigma_{x_1} < 1.5$, colour parameter $|c| < 0.3$, and colour uncertainty $\sigma_c < 0.2$, similar to the cuts of Foundation \citep{foley18}, Pantheon \citep{scolnic18}, and Pantheon+ \citep{scolnic22}.
The samples are also sculpted by the temporal coverage cut of \citet{rubin23} which requires photometry within a few days of maximum light and over a week between the first and last observation.
Outliers were detected and removed by comparing reduced $\chi^2$ values against sample-specific thresholds, by identifying points where the fitters diverged in estimating similar model parameters, and by analysing the inlier/outlier probabilities from the mixture-model in the Unified Nonlinear Inference for Type Ia cosmologY framework \citep[UNITY;][]{rubin15, rubin23, rubin26}.

The \hsf{} data release includes a third sample based on SNooPy's max\_model which we do not analyse in this study due to a cut requiring observed bandpasses to map onto rest-frame $J$, $V$, and $r$ bandpasses.\footnote{As defined by the Carnegie Supernova Project, with total transmission functions available at \url{https://csp.obs.carnegiescience.edu/data/filters}.}
This requirement effectively defines an upper limit on redshift of $z \lesssim 0.073$ in the SNPY\_Max sample, a limit which has several consequences.
First, the smaller volume of this sample would make bulk flow estimates incomparable with those based on the SNPY\_EBV or SALT samples.
The highest CMB-frame redshift in the \hsf{} dataset is that of SN 21fzq at $z_\text{obs} = 0.113294 \pm 4\times10^{-6}$, which is in both the SNPY\_EBV and SALT samples.
The median redshift of the SNPY\_Max sample is $\approx 0.0452$, about 18\% lower than the median redshifts of $\approx 0.0554$ and $\approx 0.0548$ in the SNPY\_EBV and SALT samples.
While the bulk flow inferred from the SNPY\_Max sample would still be comparable with predictions from $\Lambda$CDM, it would not serve our secondary focus of analysing the systematic consequences of choosing one methodology or sample over another.
Second, the qualitatively different selection function would require unique and more sophisticated preprocessing (to be discussed in Section \ref{sec:preprocessing}).
Additionally, the analysis in \citet{do25} indicated that the intrinsic dispersion and the root-mean-square (RMS) dispersion of Hubble Residuals within the SNPY\_Max sample were nearly a factor of 1.5 to 2 times greater than the corresponding values within the SNPY\_EBV and SALT samples.
For these reasons, we do not consider the distances in the SNPY\_Max sample for the rest of this paper.

The SNPY\_EBV and SALT samples can both be divided into two additional samples: one containing the results of applying the same survey-wide and model-specific cuts but using only optical data (sample name appended with ``\_O'' for Optical), and another using optical and NIR data but limited to targets found in the corresponding optical-only sample (appended with ``\_OJ'' for Optical and $J$-band).
The final 6 samples, brief descriptions, and constituent numbers of \sneia{} are presented in Table \ref{tab:samples}.
Each optical and optical + $J$-band pair of samples contains the same \sneia{} such that differences in analyses cannot be ascribed to sample selection and must be attributed to the addition of NIR data.

\begin{table}
    \centering
    \begin{tabular}{|c|c|c|}
    \hline
        Sample Name & Description & $N_{\text{SN}}$ \\ \hline
        SNPY\_EBV & Fit with \texttt{SNooPy}'s EBV\_model2 & 357 \\ \hline
        SNPY\_EBV\_O & Optical-only subset & \multirow{2}*{212} \\ \cline{1-2}
        SNPY\_EBV\_OJ & Optical and J-band subset &  \\ \hline
        SALT & Fit with SALT3-NIR & 362 \\ \hline
        SALT\_O & Optical-only subset & \multirow{2}*{348} \\ \cline{1-2}
        SALT\_OJ & Optical and J-band subset &  \\ \hline
    \end{tabular}
    \caption{
    The samples presented in the \hsf{} dataset are defined by the model used to infer distance moduli and whether the data used for inference include only optical or both optical and NIR data.
    The primary samples (no O(J) suffix) and the OJ samples both use optical and NIR data, but the latter have been truncated to comprise the same \sneia{} in their corresponding O samples.
    }
    \label{tab:samples}
\end{table}

The NIR observations in \hsf{} were performed with UKIRT on Maunakea, which has a declination limit of $-40\degree < \delta < 60\degree$.
Furthermore, the number density of targets in each sample is not constant throughout the survey volume, with lower Galactic latitudes at southern declinations disproportionately affected by the cut based on Galactic extinction.
This non-spherically-symmetric survey volume means inaccuracies in the calibration of distances could either produce a spurious bulk flow signal or lead to the unwarranted ``correction'' of a real bulk flow signal.
The calibration within each sample is detailed in Section 3 of \citet{do25}, but in brief, the distance moduli inferred in each sample are not based on measurements of absolute magnitude in nearby \sneia{}.
Indeed there are zero-point offsets between each sample, which will cause the peculiar velocities to differ by an additive term scaling roughly linearly with $z_{\text{cos}}$.

\subsection{Preprocessing}
\label{sec:preprocessing}
The distances within the \hsf{} dataset did not include priors accounting for the homogeneous or inhomogeneous Malmquist bias (See Appendix \ref{appendix:malmquist}).
We construct a basic Bayesian model where the given distance moduli ($\mu_{\text{obs}}$), uncertainties ($\sigma_{\mu}$), and sample-specific intrinsic dispersion ($\sigma_{\text{int}}$) are used to estimate the ``True'' distance moduli ($\mu_{\text{true}}$).
In this model, we assume the parameter values and uncertainties in the \hsf{} dataset are normally distributed, such that
\begin{equation}
    \mu_{\text{obs},n}|\mu_{\text{true},n} \sim \mathcal{N}(\mu_{\text{true},n},\sigma_{\mu,n}^2 + \sigma_{\text{int}}^2)
\end{equation}
where $\sigma_{\text{int}}$ is provided for each sample as the intrinsic dispersion.

A more accurate and robust model would require accounting for the selection function,\footnote{Section 8.1 of \textit{Bayesian Data Analysis} \citep[3rd ed.][]{gelman04} is titled ``Bayesian inference requires a model for data collection''.} but it is notoriously difficult to construct a principled analytic formulation.
In the UNITY framework \citet{rubin15, rubin23, rubin26} selection is semi-analytically modelled with survey-specific limiting magnitudes and error functions.
The Bayesian Estimation Applied to Multiple Species \citep[BEAMS;][]{kunz07} with Bias Corrections \citep[BBC;][]{kessler17} method uses realistic simulations to infer a non-analytic selection efficiency across multiple dimensions, provisioning corrections that bring the observed sample in line with the expected population.
A recent simulation-based inference (SBI) approach uses normalising flows to infer a non-analytic likelihood for selection, and initial results show an increase in robustness against the assumption of a fiducial cosmology seen in the BBC approach \citep{boyd24, boyd26}.
As an alternative approach, uncertainty due to selection effects can be mitigated if not avoided by restricting the sample of observed \sneia{} to regions of full completeness as done in the Zwicky Transient Facility (ZTF) volume-limited sample \citep{amenouche25, ginolin25b}.

For this study, we use a phenomenological selection similar to the one introduced by \citet{lavaux16}, but with a selection cut-off parametrized by an error-function complement rather than an exponential decay:
\begin{equation}
    \pi(r) = Cr^2~\text{erfc}\left( \frac{r - R_{\text{cut}}}{\sigma_{\text{cut}}} \right)
    \label{eqn:r_prior}
\end{equation}
where $C$ is a normalisation constant such that $\int_0^\infty\pi(r)\, \mathrm{d}r = 1$, and $R_{\text{cut}}$ and $\sigma_{\text{cut}}$ are jointly inferred for each sample using the following hyper-priors: the characteristic distance of the cut-off is in the upper half of observed distances $R_{\text{cut}} \sim \mathcal{U}(\text{median}(\vec{r}_{\text{obs}}),\text{max}(\vec{r}_{\text{obs}}))$ Mpc, and its scale factor is positive and less than the largest observed distance $\sigma_{\text{cut}} \sim \mathcal{U}(0,\text{max}(\vec{r}_{\text{obs}}))$ Mpc.
The normalisation constant $C$ is calculated in Appendix \ref{appendix:selection_norm}, with the derivation only valid if the power-law index is an integer.
For this reason, combined with the expectation that \sneia{} have a constant comoving number density, we fix the power-law index at $2$ rather than sample a continuous distribution of values.

This phenomenological selection function applied to distance moduli has shortcomings.
It does not address Malmquist bias \citep[][not to be confused with the homogeneous Malmquist bias described in Appendix \ref{appendix:malmquist}]{malmquist22}, where some of the targets near the detection limit are likely to have scattered \textit{inwards} from regions beyond the nominal survey volume.
On the contrary, targets near the detection limit are interpreted as having scattered \textit{outwards} from regions where the distance prior indicates we are more likely to detect targets \citep[compensating for ``Eddington Bias'',][]{eddington13}.
Properly accounting for Malmquist bias would require working with the photometry of the dataset, which is a layer deeper than our analysis of distance moduli.
Another potential issue is the sensitivity of the inferred latent distance moduli to the hyper-priors on $R_{\text{cut}}$ and $\sigma_{\text{cut}}$, which are defined arbitrarily.
Also, the function converting between proper distances and distance moduli requires values for $H_0$ and $\Om$, which provides a vector for inaccurate values to bias the inferred distance moduli.
We use the $H_0$ values quoted within the \hsf{} data files and since those values are closer to the SH0ES $H_0$ than the Planck $H_0$, we use the SH0ES value for $q_0 = -0.51$ \citep{riess22}.

The Hamiltonian Monte Carlo inference is performed using \texttt{numpyro} \citep{numpyro, pyro} and its No-U-turn sampler.
We assess convergence by verifying that the split Gelman-Rubin scale reduction factor $\widehat{R}$ \citep{gelman92, vehtari21} is at most 1.05.
This led us to use 1000 warm up steps and draw 2500 samples in each of four chains.
The end result of pre-processing is a set of ``true'' distance moduli $\mu_{\text{preproc}}$ and uncertainties based on the medians and standard deviations of the posterior samples.
\section{Validating Methodologies}
\label{sec:validation}

\citet{whitford23} stress that the idiosyncratic interactions between bulk flow estimators and various surveys are best characterized by performing analyses on realistic mock catalogues.
The level of realism in a simulation is difficult to quantify, but the end points are relatively well-defined.
The most simplistic mock catalogue is composed of randomly and independently distributed position and velocity vectors, while the most realistic mock catalogues incorporate the linear and non-linear aspects of structure growth using  N-body simulations performed in a theoretically motivated cosmology and account for observational and selection effects.

\subsection{Simulated Catalogues}
We generate 8 sets of mock catalogues to span three binary decisions:
\begin{itemize}
    \item is the simulated peculiar velocity field uniform or induced by a point-source?
    \item is the catalogue spherically symmetric or declination-limited?
    \item are the cosmological parameters assumed from \citet{planck18} or  \citet{riess22}?
\end{itemize}
Each point will be addressed in its corresponding subsection, but we will first describe features common to all catalogues.

We simulate 300 tracers in all catalogues, distributing them uniformly within a comoving survey volume of radius $R_{\text{survey}} = 500$ Mpc.
The covariance matrix for the radial peculiar velocities
\begin{equation}
    G_{mn} = \delta^K_{mn}\sigma_\text{NL}^2  + \frac{H_0^2 f^2}{2\pi^2} \int_0^\infty P(k) f_{mn}(k) \mathrm{d}k
    \label{eqn:pv_covar}
\end{equation}
where $\delta^K_{mn}$ is the Kronecker delta, $\sigma_{\text{NL}}$ is the uncertainty in radial velocity due to non-linear effects, and $\bm{f}(k)$ is the correlated window function from Equation \ref{eqn:corr_window_fn}.\footnote{The diagonal terms are undefined under the spherical harmonic decomposition \citep[Equation A11;][]{ma11}, but the original integral (Equation A1) becomes simple without the exponential term: $f_{mm}(k) =\int \frac{\mathrm{d}\hat{k}}{4\pi} \left(\hat{r}_m \cdot \hat{k}\right)^2 = 1/3$.}
The covariance matrix requires cosmological parameters and a matter power spectrum, which we generate with CAMB \citep{lewis00, howlett12}.
Each mock catalogue assigns radial velocities drawn from a multivariate normal distribution using the covariance matrix and a mean based on $u(\vec{r})$, which takes one of the two forms detailed in Section \ref{sec:validation_varying_pv_source}.

In all sets of catalogues, we simulate observational uncertainties by adding a random draw from $\mathcal{N}(0,\sigma_\text{HSF}^2)$ to each simulated distance modulus or observed redshift.
The variance $\sigma_\text{HSF}^2$ is the median variance in distance modulus or observed redshift of the 10 targets in the SNPY\_EBV catalogue with CMB-frame redshifts closest to the object's simulated cosmological redshift.
This is intended to replicate the difference in precision between distance modulus estimates of near and far SNe.

None of our mock catalogues simulate the selection effects one expects near the detection limit.
The homogeneous Malmquist bias associated with uniform comoving number density is simulated by generating uniformly distributed comoving coordinates for each target and adding a symmetric error term to the distance moduli.
We do not simulate LSS, meaning the catalogues are not affected by the inhomogeneous Malmquist bias.
Non-linear effects are modelled as an additional layer of uncertainty introduced by the $\delta^K_{mn}\sigma_\text{NL}^2$ term in the covariance matrix.
All catalogues are generated with $\sigma_\text{NL} = 300$ km s$^{-1}$.

Our coordinate frame is oriented with $\hat{x}^x$ pointing towards $(\alpha,~\delta)$ = (0\degree, 0\degree), $\hat{x}^y$ towards (90\degree, 0\degree), and $\hat{x}^z$ towards $\delta=90\degree$.

\subsubsection{Varying Peculiar Velocity Source}
\label{sec:validation_varying_pv_source}
The first decision point divides our catalogues into those with uniform peculiar velocity fields and those with fields induced by a point-source.

The uniform peculiar velocity fields ($\vec{v}(\vec{r}) = \vec{B}_0$ and $u(\vec{r}) = \vec{B}_0 \cdot \hat{r}$) are oriented randomly over the unit-sphere and have speeds drawn from a uniform distribution $\mathcal{U}(0,2000)$ km s$^{-1}$.
This is the signal one might expect in a ``dipole'' or ``tilted'' cosmology \citep[e.g.,][]{krishnan23, ebrahimian24}, but not what one would expect from the tidal forces induced by a large over/under-density beyond the survey volume.

The point-source-induced peculiar velocity fields are simulated by drawing a mass $m_\text{sim} \sim \mathcal{U}(-m_{\text{lim}},m_{\text{lim}})$ located at $\vec{r}_{\text{sim}}$ where $|\vec{r}_{\text{sim}}| \sim \left[r^3_{\text{min}}+(r^3_{\text{max}} - r^3_{\text{min}})\mathcal{U}(0,1)\right]^{1/3}$ such that $\vec{r}_{\text{sim}}$ is uniformly distributed in volume between $r_{\text{min}} = R_{\text{survey}}$ and $r_{\text{max}} = 2R_{\text{survey}}$.
The mass range is such that a mass of $-m_{\text{lim}}$ exactly counters the mass of matter within a ball of radius $r_{\text{sim}}-R_{\text{survey}}$ and uniform density $\rho_\text{M} = \Om\rho_{\text{crit}} = \Om\frac{3H_0^2}{8\pi G}$, creating an uncompensated void with no gravitating matter.
The upper-limit is arbitrarily defined to replicate an uncompensated overdensity with average $\rho = (1+\Om)\rho_{\text{crit}}$ and radius $r_{\text{sim}}-R_{\text{survey}}$.
The limits on $r_{\text{sim}}$ are chosen such that even if $m_{\text{sim}}$ takes an extreme value of $-m_{\text{lim}}$, the equivalent void could still be fully outside the survey volume, justifying its characterisation as a point-source.
Furthermore, this leaves the matter power spectrum used for generating the radial peculiar velocity covariance matrix unchanged if evaluated strictly within the survey volume.
That said, the addition/subtraction of mass means the matter power spectrum no longer describes the totality of gravitational forces within the survey volume, and a more sophisticated study should either include a perturbation term in the power spectrum, adjust the limits of integration in $k$ space, or use N-body simulations to look at survey volumes neighbouring voids or clusters to study tidal peculiar velocity fields.
The radial projection of the velocity field adds the following term to each object in the catalogue \citep[modified from Equation 18.1.10 of][]{coles02}:
\begin{equation}
    u(\vec{r}) = \vec{v}(\vec{r})\cdot \hat{r} = - \frac{2 G f m_{\text{sim}}}{3\Om H_0} \frac{(\vec{r} - \vec{r}_{\text{sim}})}{|\vec{r} - \vec{r}_{\text{sim}}|^3} \cdot \hat{r}.
    \label{eqn:rad_pv_pointsource}
\end{equation}
Using a uniform weighting function in Equation \ref{eqn:bulk_alt}, the bulk flow over a survey volume $V$ projected to Cartesian unit vector $\hat{x}^{\alpha}$ is
\begin{equation}
    \vec{B} \cdot \hat{x}^{\alpha} = -\frac{2 G f m_\text{sim}}{H_0 \Om V}\int_V \mathrm{d}V r \frac{(\vec{r} - \vec{r}_{\text{sim}}) \cdot \hat{r}}{|\vec{r} - \vec{r}_{\text{sim}}|^3} \hat{r} \cdot \hat{x}^\alpha,
    \label{eqn:tidal_bulk_flow}
\end{equation}
which is equivalent to $\vec{v}(\vec{0}) \cdot \hat{x}^{\alpha}$ for a spherically symmetric survey volume.

\subsubsection{Varying Survey Geometry}
\label{sec:validation_varying_survey_geometry}
The next decision point addresses the angular selection function: whether targets are generated over a spherical survey volume or within a declination limit of $-40 \degree < \delta < 60\degree$, matching the \hsf{} samples.
The declination limit leads to several effects:
first, since the number of targets in each catalogue are held constant, the decrease in survey volume increases the spatial density of targets;
second, the window function loses isotropy and peculiar velocities aligned with the declination-axis become more difficult to constrain; third, the lack of spherical symmetry in the survey volume means that the bulk flows induced by point-sources are not equal to  $\vec{v}(\vec{0})$ and are instead calculated using the method described in Appendix \ref{appendix:bulk_flow_integral}. 

\subsubsection{Varying Cosmological Parameters}
\label{sec:validation_varying_cosmo}
The final decision point covered in this set of catalogues deals with the cosmological parameters assumed in each simulation.
Given the computational cost of generating and analysing simulations, we decided against varying cosmological parameters continuously and opted to test two distinct sets of values, those from the ``Plik best fit'' of \citet{planck18} ($H_0=67.3117$ km s$^{-1}$ Mpc$^{-1}$ and $\Om \approx 0.3158$), and those from the joint $H_0 - q_0$ fit of \citet{riess22} ($H_0=73.30$ km s$^{-1}$ Mpc$^{-1}$ and $\Om \approx 0.3267$).

The cosmological parameters $H_0$ and $\Om$ appear directly in the radial peculiar velocity covariance matrix (Equation \ref{eqn:pv_covar}, $\Om$ appearing through $f \approx \Om^{0.55}$), and indirectly through $P(k)$.
We generate a matter power spectrum for the Planck set of parameters using the CAMB-provided ini file.\footnote{\url{https://raw.githubusercontent.com/cmbant/CAMB/master/inifiles/planck_2018.ini}}
The spectrum for the SH0ES values of $H_0$ and $\Om$ are initialised from the same file, but $H_0$ is modified to be 73.3 km s$^{-1}$ Mpc$^{-1}$ and the baryon, cold dark matter, and massive neutrino density fractions $\Omega_b h^2$, $\Omega_c h^2$, and $\Omega_\nu h^2$ are all increased by about 23\% such that $\Om = 0.3267$.
Both power spectra are generated at redshift $z=0$ and scaled to units of Mpc$^3$ rather than $h^{-3}$ Mpc$^3$ to be consistent with the units of our simulated coordinates.

The different cosmological parameters are also relevant to the point-source-induced peculiar velocity fields through Equation \ref{eqn:tidal_bulk_flow} and indirectly through $m_\text{lim} \propto \Om H_0^2$.


\subsection{Simulation Results}
We analyse each mock catalogue four times: twice using the Nusser MLE method and twice using the Peery MVE method, duplicating analyses to vary the cosmology assumed in the analysis code (similar to, but distinct from the variation in Section \ref{sec:validation_varying_cosmo}).
This strategy allows us to evaluate the methods when the assumed cosmological parameters exactly match the data-generating process and when they differ.

The Nusser MLE method includes a term for the spatial density of tracers as a function of distance.
This term is not related to clustering, population-drift over redshift, or anything astrophysical, but is ``entirely due to observational selection strategy'' \citep{nusser14}.
Our simulations use a constant spatial density (ignoring the survey edges), so we provide an arbitrary density of 1 tracer per cubic Mpc for all tracers in all catalogues.

\begin{figure*}
    \includegraphics[width=\textwidth]{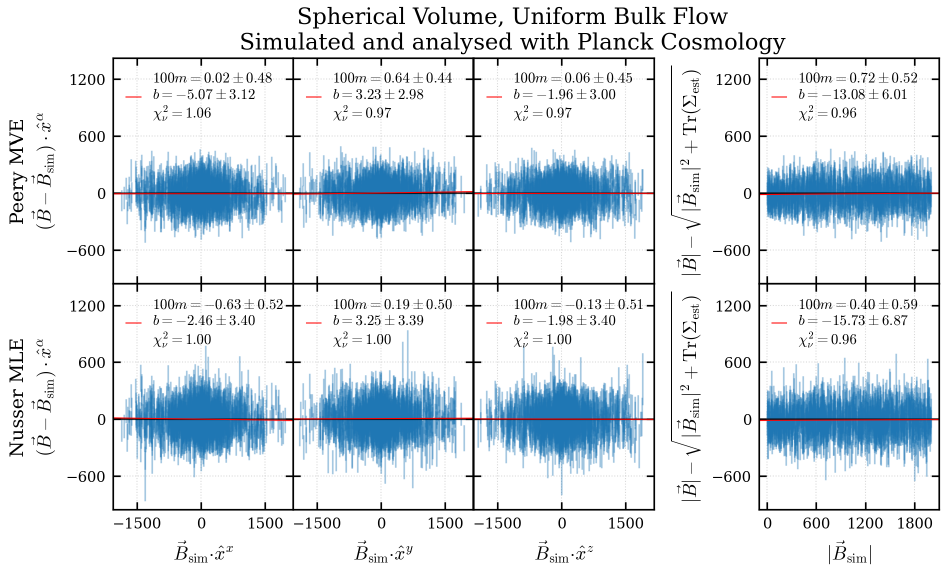}
    \caption{We use the Peery MVE (top row) and Nusser MLE (bottom row) methods to analyse 1500 mock catalogues with simulated bulk flows.
    The differences between the recovered ($\vec{B}$) and simulated ($\vec{B}_\text{sim}$) values are plotted against simulated values, with the left three columns representing Cartesian projections of the bulk flow and the rightmost column representing the total speed of the bulk flow.
    The differences in the rightmost column include the trace of the covariance matrix of $\vec{B}$ to account for the positive expectation values of Maxwellian distributions.
    The red lines show the best-fitting linear models ($y = mx+b$) with inverse-variance weighting.
    All quantities are in units of km s$^{-1}$ except for the slopes $m$ and $\chi^2_\nu$ values which are dimensionless.
    Note that the presented slope values are scaled by a factor of 100.
    The uncertainties on $|\vec{B}|$ were calculated with the covariance matrix, which may differ from the root quadrature sum of uncertainties on $\vec{B}\cdot \hat{x}^\alpha$.
    The bulk flows presented here were simulated as a uniform peculiar velocity field over a fully spherically symmetric survey volume, assuming a Planck cosmology during simulation and analysis.
    }
    \label{fig:mocks_symm_uniform_planck_planck}
\end{figure*}

We create plots showing the residuals between recovered and simulated values for the Cartesian projections and total speeds of our simulations.
These latter differences are not simple subtractions of the recovered and simulated speeds because there is a strictly positive contribution from cosmic variance, represented as the trace of the recovered bulk flow covariance matrix $\bm{\Sigma}_{\text{est}}$.
Each plot includes the best-fitting inverse-variance-weighted linear models between the simulated values and residuals, for which we expect an unbiased recovery to have a negligible slope ($m$) and intercept ($b$), and a $\chi^2_\nu$ near 1.
A representative plot is shown in Figure \ref{fig:mocks_symm_uniform_planck_planck}, which demonstrates unbiased recovery of the simulated bulk flow.
Not all analyses were as successful, and in the following sub-sections, we present issues identified during testing.

\subsubsection{Definitional Discrepancy}
\label{sec:validation_definitional_differences}
\begin{figure*}
    \includegraphics[width=\textwidth]{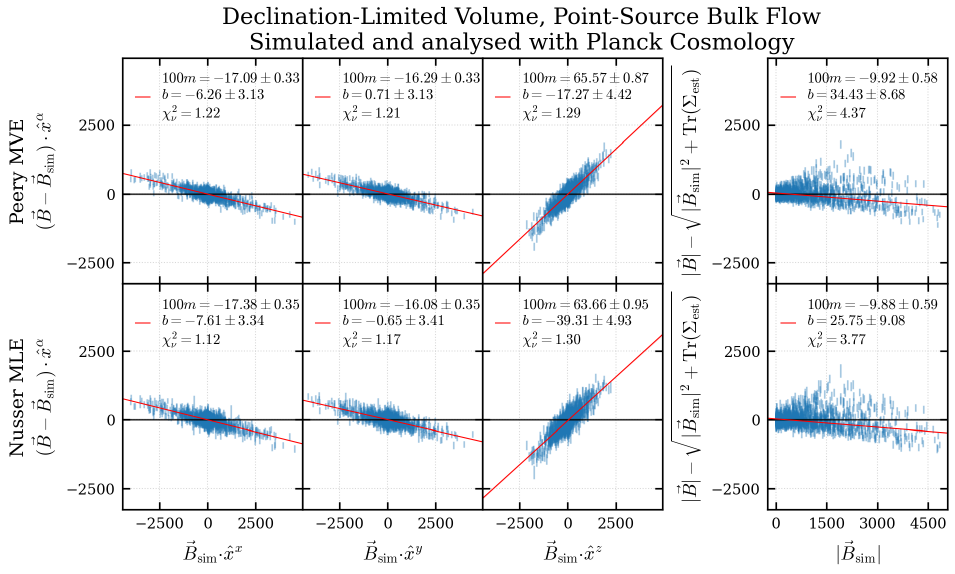}
    \caption{The same format as Figure \ref{fig:mocks_symm_uniform_planck_planck}, but the bulk flows presented here were simulated as the field induced by a point-source mass outside a declination-limited ($-40 \degree < \delta < 60 \degree$) survey volume. The $\vec{B}_\text{sim}$ values were calculated for the actual survey volume, and the significant slopes are evidence of the definitional discrepancy described in Section \ref{sec:validation_definitional_differences}.
    }
    \label{fig:mocks_declim_pointsource_planck_planck}
\end{figure*}

\begin{figure*}
    \includegraphics[width=\textwidth]{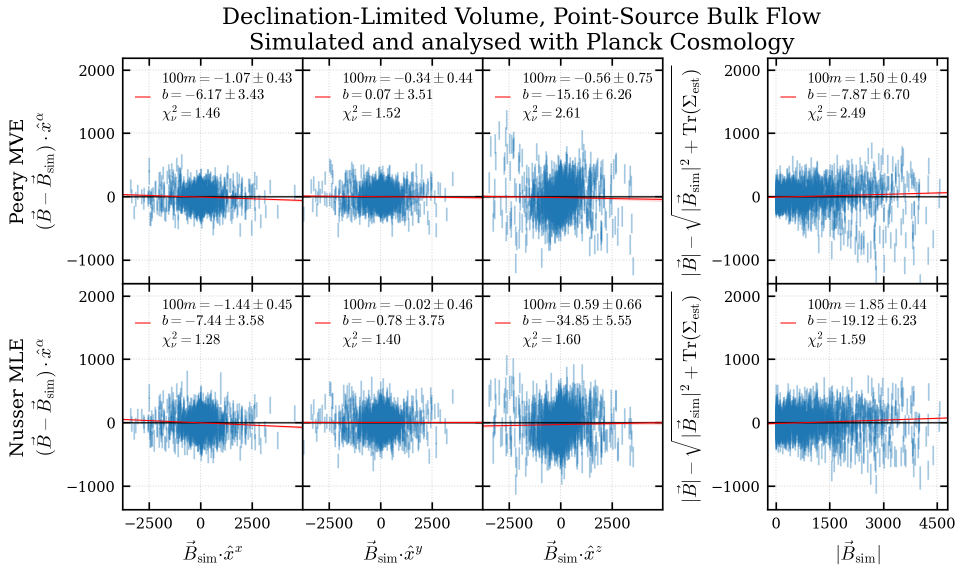}
    \caption{The same format as Figure \ref{fig:mocks_declim_pointsource_planck_planck}, but the definitional discrepancy described in Section \ref{sec:validation_definitional_differences} is avoided by instantiating the tracers within a declination-limited volume but calculating $\vec{B}_\text{sim}$ as though in a spherically symmetric volume.
    }
    \label{fig:mocks_declim_pointsource_planck_planck_symm}
\end{figure*}

The bulk flow induced by a point-source mass in a declination-limited survey volume is not accurately inferred by either method, with Figure \ref{fig:mocks_declim_pointsource_planck_planck} showing how the axisymmetric volume leads to underestimated speeds ($m\approx 15\%$) when projected to the $x$ and $y$ Cartesian axes and overestimated speeds ($m\approx 60\%$) when projected to the $z$-axis.
This is because the simulated values were calculated for the declination-limited survey volume while the recovery methods were designed to infer the bulk flow over a spherically symmetric volume.
The Nusser MLE returns estimates of the bulk flow as defined from Equation 1 of \citet{nusser14}, which describes the bulk flow within a spherical volume, not an arbitrary one.
Similarly, the MVE \citep{watkins09} and Peery MVE \citep{peery18} methods first imagine an idealised survey with tracers distributed uniformly over the unit sphere, then they calculate weights for the observed tracers that minimise the variance between the bulk flows inferred by the imagined and actual catalogues.
Figure \ref{fig:mocks_declim_pointsource_planck_planck_symm} shows the same inferences from the Nusser MLE and Peery MVE methods, but plotted against the bulk flow expected for a spherically symmetric survey volume.
This figure shows less biased recovery of the simulated bulk flow, although the $z$-axis residuals from both methods show some structure with extreme bulk flow moments tending to be misestimated.
This issue extends to the total bulk flow speeds.
When simulating bulk flows with low speeds, the methods seem to recover relatively unbiased, unimodal estimates, but at higher simulated speeds, the recoveries become multimodal.
From this test, we first conclude that the recovered bulk flows should be compared against simulated values calculated for spherically symmetric survey volumes regardless of the simulated survey's footprint.
We also conclude that while the recovered bulk flow may be unbiased ($|m|\lesssim 2\%$) on average, an inferred speed may correspond to a multimodal distribution of ``true'' speeds.
Given the contrived nature of the simulations (one point-like gravitational source), any probabilistic model for inferring a true speed based on a measured one constructed from these results will only be appropriate for similarly contrived simulations.
However, the significant $\chi^2_\nu$ suggests that the statistical variance should be increased by a systematic term when analysing a declination-limited survey and assuming a non-uniform peculiar velocity field.

\subsubsection{Cosmological Differences}
\label{sec:validation_cosmology_differences}
\begin{figure*}
    \includegraphics[width=\textwidth]{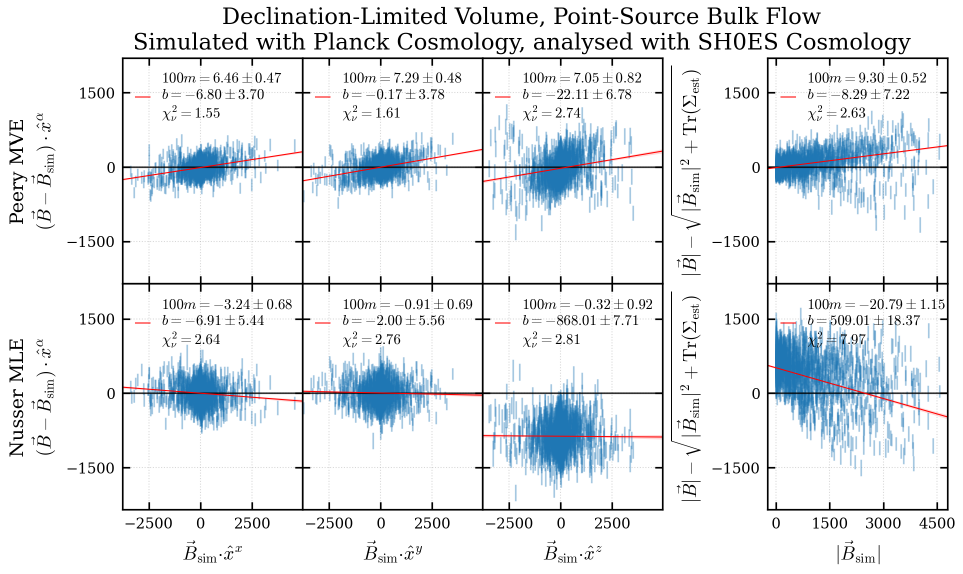}
    \caption{The same format as Figure \ref{fig:mocks_declim_pointsource_planck_planck_symm}, but the recovery methods assumed a SH0ES cosmology, deviating from the Planck cosmology used in the simulation.}
    \label{fig:mocks_declim_pointsource_planck_SH0ES_symm}
\end{figure*}

As mentioned in Section \ref{sec:measuring_bfs}, using inaccurate values of the cosmological parameters to infer peculiar velocities will bias all measurements by a monopole term varying with distance only ($\Delta u(r) = u_{\text{obs}}(r) - u_{\text{sim}}(r)$).
The effects of a monopole may cancel out in spherically symmetric surveys, but will otherwise leak into the inferred bulk flow projections.
This is seen in Figure \ref{fig:mocks_declim_pointsource_planck_SH0ES_symm}, in which data were simulated with a Planck cosmology, but analysed with a SH0ES cosmology.
The Peery MVE method includes explicit constraints to mitigate the consequences of this mismatch, but the recovered Cartesian projections appear overestimated by $\approx 6\%$ and the total speed by $\approx 9\%$.
The Nusser MLE method is more severely affected, and also shows a $\approx 840$ km s$^{-1}$ offset in the recovered bulk flow projected along the $z$-axis due to the monopole leakage.

\section{Results}
\label{sec:results}

The bulk flow estimates from the Nusser MLE and Peery MVE methods applied to the SNPY\_EBV and SALT samples and their ``O'' and ``OJ'' variants are provided in Table \ref{tab:bf_vels}.

\begin{table*}
    \centering
    \begin{tabular}{|r|c|c|c|c|c|c|c|}
        \hline
        Sample & Estimator & $\vec{B}\cdot \hat{x}^x$ & $\vec{B}\cdot \hat{x}^y$ & $\vec{B}\cdot \hat{x}^z$ & $|\vec{B}|$ & $\sqrt{\text{Tr}(\bm{\Sigma}_{\text{cos}})}$ & p-value \\
        \hline
        SNPY\_EBV & Peery MVE & -11.6(112.7) & -11.3(130.8) & -240.7(173.6) & 241.2(174.2) & 162.4 & 0.709\\
        SNPY\_EBV & Nusser MLE & -104.9(128.7) & 31.2(139.4) & -194.7(190.4) & 223.4(174.1) & 210.2 & 0.757\\
        \hline
        SNPY\_EBV\_O & Peery MVE & 127.2(153.3) & 44.6(166.7) & -305.5(242.5) & 333.9(222.2) & 176.7 & 0.544\\
        SNPY\_EBV\_O & Nusser MLE & 37.3(166.8) & 35.2(178.2) & -209.7(235.3) & 215.9(224.9) & 212.3 & 0.869\\
        \hline
        SNPY\_EBV\_OJ & Peery MVE & 104.0(148.0) & -49.6(160.6) & -374.1(231.9) & 391.4(223.1) & 172.5 & 0.450\\
        SNPY\_EBV\_OJ & Nusser MLE & 30.5(160.6) & -30.3(173.5) & -238.3(226.4) & 242.1(223.7) & 209.6 & 0.832\\
        \hline
        SALT & Peery MVE & 30.1(112.7) & -48.3(138.7) & -196.6(185.4) & 204.6(185.1) & 166.4 & 0.822\\
        SALT & Nusser MLE & -60.5(246.4) & -110.2(168.3) & -291.5(299.1) & 317.5(285.2) & 322.6 & 0.808\\
        \hline
        SALT\_O & Peery MVE & 25.8(119.1) & -42.1(147.0) & -88.9(201.6) & 101.7(195.7) & 172.1 & 0.975\\
        SALT\_O & Nusser MLE & -15.4(199.9) & -116.4(181.8) & -139.8(314.9) & 182.6(280.9) & 310.0 & 0.936\\
        \hline
        SALT\_OJ & Peery MVE & 14.0(113.6) & -56.1(138.5) & -162.4(187.2) & 172.4(186.8) & 167.4 & 0.889\\
        SALT\_OJ & Nusser MLE & -34.3(196.8) & -94.5(166.7) & -255.1(302.8) & 274.2(290.6) & 308.5 & 0.862\\
        \hline
    \end{tabular}
    \caption{We use the Peery MVE and Nusser MLE methods to estimate bulk flow in the SNPY\_EBV, SNPY\_EBV\_O, SNPY\_EBV\_OJ, SALT, SALT\_O, and SALT\_OJ samples.
    We summarise the cosmic covariance matrices $\bm{\Sigma}_{\text{cos}}$ with the root of their traces, but use the full matrices in combination with the estimator covariance matrices when calculating p-values (See Equation \ref{eqn:bf_p_value}).
    Parenthetical values are 1$\sigma$ uncertainties, and all quantities are in units of km s$^{-1}$ except for the p-values, which are dimensionless.}
    \label{tab:bf_vels}
\end{table*}

\subsection{Comparisons with Theory}
Section \ref{sec:lcdm} describes the expected consequences of basic assumptions (cosmological principle, FLRW metric, $\Omega_K = 0$, linear peculiar velocity dominates over non-linear effects at scales above $\sim 10$ Mpc): the Cartesian projections of bulk flow should be distributed as a multivariate normal $\vec{B} \sim \mathcal{N}(\vec{0},\bm{\Sigma}_{\text{cos}})$, where $\bm{\Sigma}_{\text{cos}}$ is calculated from Equation \ref{eqn:lcdm_pv_variance} using the sample-specific window-functions $\bm{\tilde{W}}^2(k)$ from Equation \ref{eqn:data_window_function}.
This angular flexibility is important because a declination-limited survey like Hawai`i Supernova Flows has reduced leverage for constraining bulk flows aligned with the $z$-axis over large spatial scales, leading to more significant cosmic variance in the inferred bulk flow projection $\vec{B} \cdot \hat{x}^z$.
To provide a sense of scale without adding 2 or 8 more columns to Table \ref{tab:bf_vels} we summarise these covariance matrices using a scalar, but we use the full matrices in combination with the estimator covariance matrices when calculating p-values.

\begin{figure}
    \centering
    \includegraphics[width=\linewidth]{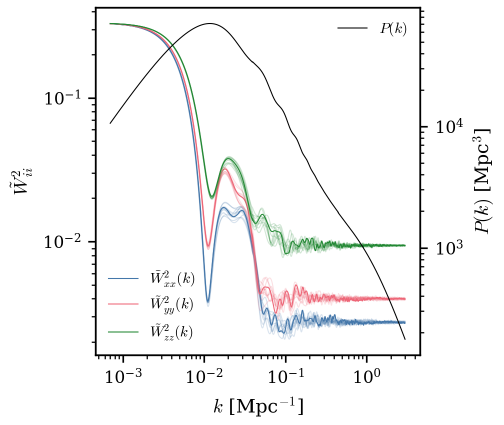}
    \caption{We present the diagonal components of the tensor angle-averaged window-function $\bm{\tilde{W}}^2(k)$ for the SNPY\_EBV sample, calculated using Equation \ref{eqn:data_window_function} and weights inferred for the Peery MVE.
    The matter power spectrum generated with CAMB using $H_0 \approx 70.7$ km s$^{-1}$ Mpc$^{-1}$ and $q_0=-0.51$ is provided for comparison since $\Sigma_{ij,\text{cos}} \propto \int_0^\infty P(k) \tilde{W}_{ij}^2(k)\mathrm{d}k$.
    The solid lines show the nominal window functions calculated with $\mu_\text{obs}$, and in reduced opacity, we plot 10 additional window functions calculated with the same weights but with random realizations of $\mu_\text{obs}$ based on $\sigma_\mu$.
    }
    \label{fig:window_fn}
\end{figure}

The diagonal components of the window function calculated for the SNPY\_EBV sample using weights from the Peery MVE is shown in Figure \ref{fig:window_fn} as an example.
The nominal window function components are calculated using $\mu_\text{obs}$ values taken from the median of the pre-processing posterior samples.
Variation due to uncertainties in distance moduli is presented through alternative window functions calculated with $\bm{f}(k)$ functions based on random realizations $\mu_\text{alt} \sim \mathcal{N}(\mu_\text{obs},\sigma^2_\mu)$.
This causes individual distances to vary by 5\% to 10\%, which affects the higher spatial-frequencies ($k \gtrsim 10^{-2}$ Mpc$^{-1}$) much more than the lower ones.
However, the log-scale in $k$ of Figure \ref{fig:window_fn} disguises the relatively greater contribution of higher frequencies when integrating $\int_0^\infty P(k)\tilde{W}^2_{ij}(k)\mathrm{d}k$.
Thus, although the integrand is an order of magnitude greater where uncertainties in distance do not affect the window function, the effect on the covariance matrix $\bm{\Sigma}_{\text{cos}}$ are not completely negligible.
The standard deviations of the variances are about 1\% of their nominal values, while the standard deviations of the covariances are around 10\% to 20\%.

We test the null hypothesis that $\vec{B} \sim \mathcal{N}(\vec{0},\bm{\Sigma}_{\text{cos}})$ when $\vec{B}$ is inferred at $\vec{B}_\text{est}$ with covariance matrix $\bm{\Sigma}_{\text{est}}$.
The squared Mahalanobis distance \citep{mahalanobis30} parametrized by $\vec{B}_\text{est}$ and $\bm{\Sigma}_\text{cos} + \bm{\Sigma}_\text{est}$ should follow a $\chi^2$-distribution with 3 degrees of freedom \citep[Chapter 4.2;][]{johnson07}.
The probability of measuring a bulk flow with velocity and covariance matrix more extreme than $\vec{B}_\text{est}$ and $\bm{\Sigma}_{\text{est}}$ is calculated as
\begin{equation}
    p = 1-\frac{\gamma\left(\frac{3}{2}, \frac{1}{2}\vec{B}_\text{est}^T (\bm{\Sigma}_{\text{cos}} + \bm{\Sigma}_{\text{est}})^{-1} \vec{B}_\text{est} \right)}{\Gamma(3/2)}
    \label{eqn:bf_p_value}
\end{equation}
where $\gamma$ is the lower incomplete Gamma function \citep[\href{https://dlmf.nist.gov/8.2}{Equation 8.2.1};][]{olver10}.

Notably, these p-values are not treated for the multiple comparisons problem.
Although no sample can reject the null hypothesis that $\vec{B} \sim \mathcal{N}(\vec{0},\bm{\Sigma}_{\text{cos}})$ at even a $p \leq 0.05$ level, the shortcoming is more severe than the table indicates.
Furthermore, the significant $\chi^2_\nu$ for the simulations using a declination-limited volume and point-source bulk flow (Figure \ref{fig:mocks_declim_pointsource_planck_planck_symm}) could be easily used to justify the inclusion of a systematic error term in Equation \ref{eqn:bf_p_value} or a scaling of $\bm{\Sigma}_{\text{est}}$, either case leading to a decrease in the statistical significance of our results.
For these reasons, the p-values in Table \ref{tab:bf_vels} are best understood as lower-limits.

\subsection{Comparisons Between Estimators}
As an alternative to comparing inferred values with expectations from $\Lambda$CDM, values inferred using different estimators on the same sample may be compared against each other.
Our simulations suggest the two estimators should converge towards similar values when applied to the same data unless the cosmology assumed for analysis differs from the data-generating process.
Figure \ref{fig:mocks_declim_pointsource_planck_SH0ES_symm} shows that an overestimated value for $H_0$ will cause the Nusser MLE method to infer uniformly decreased velocities in $\vec{B}\cdot\hat{x}^z$ due to the monopole leakage in an asymmetric survey biased towards the North.
The Nusser MLE favours greater values for $\vec{B}\cdot\hat{x}^z$ than the Peery MVE for the three SNPY\_EBV samples, but more negative values for the three SALT samples.
This suggests that the $H_0$ values provided in the HSF data files are too low for the SNPY\_EBV samples (about $70.7$, $71.1$, and $70.5$ km s$^{-1}$ Mpc$^{-1}$ for the total, O, and OJ samples respectively), and too high for the SALT samples ($73.4$, $73.5$, $73.4$ km s$^{-1}$ Mpc$^{-1}$).

However, the absolute differences in $\vec{B}\cdot \hat{x}^z$ are on average about a fifth of the joint uncertainty assuming perfect correlation, or a fourth if assuming independence, and there are comparable differences in the $\vec{B}\cdot\hat{x}^x$ and $\vec{B}\cdot\hat{x}^y$ which should not suffer from a hypothetical monopole leak.
The multiplicity of the pattern should not be used to decrease statistical error without first identifying the exact nature of the expected correlation between samples due to overlapping targets and shared methodologies for reducing data from pixels to bulk flow estimates.
Such work is far beyond the scope of this paper, so we conclude only that the observation of marginally discrepant $\vec{B}\cdot \hat{x}^z$ values is suggestive, but lacking in statistical power.

\subsection{Comparisons Between Optical-only and Optical+NIR Samples}
In an analogous comparison, corresponding ``O'' and ``OJ'' samples explicitly comprise the same SNe~Ia such that differences in results may be ascribed exclusively to the inclusion of $J$-band data.
The effect of NIR photometry on distance measurements in the Hawai`i Supernova Flows data is examined Section 6.3 of \citet{do25}, which closes with the conclusion ``we cannot definitively say NIR photometry leads to smaller Hubble residuals.''
Following suit, we cannot definitively say NIR photometry leads to more accurate or more precise bulk flow inferences.
The values of $\vec{B}\cdot \hat{x}^i$ tend to be lower in the OJ samples, except for $\vec{B}\cdot \hat{x}^y$ in the SALT\_OJ sample analysed with the Nusser MLE.
These differences are on average 10\% or 15\% of the joint uncertainty assuming correlation or independence respectively.
These decreases in projected velocities produce increases in total speed of 6 to 18\% of the joint uncertainty assuming correlation or 8 to 26\% assuming independence.
Although the uncertainty in $\vec{B}\cdot \hat{x}^i$ is always lower in the OJ samples, the difference is marginal and does not extend to the uncertainties in $|\vec{B}|$ which are larger for SNPY\_EBV\_OJ analysed with the Peery MVE and SALT\_OJ analysed with the Nusser MLE.
The same caveat about multiplicity from the previous section applies here.

\section{Discussion and Conclusion}
\label{sec:discussion}

We have analysed SN~Ia-based distances and spectroscopic redshifts from the \hsf{} dataset with two well-studied methods for estimating bulk flows: the Peery MVE \citep{peery18} and the Nusser MLE \citep{nusser14}.
Across 6 defined sub-samples, there were no bulk flow inferences incompatible with predictions from linear theory.
We found no statistically significant differences in the accuracy or precision of bulk flows inferred when the distance moduli were based on optical data only or on optical and NIR data.

\subsection{Comparison to Other Work}
Although it would appear that our work agrees with $\Lambda$CDM, joining other studies finding agreement while contradicting studies finding anomalous bulk flows, the variety of methods used to estimate bulk flows based on various data sets necessitates scrutiny before directly comparing results.

It is common to present new bulk flow projections or total speeds alongside values from other analyses, but these are only indirectly comparable.
Unique datasets produce unique covariance matrices such that the expectations from theory are not identical between studies.
This can be partially remedied by reducing the geometric characteristics of a survey to an effective survey depth, and then calculating expectations with an isotropic window function like the spherical top-hat or Gaussian functions in Equations \ref{eqn:tophat_window} or \ref{eqn:gaussian_window}.
However, these general approximations do not provide the specificity needed to perform the comparisons they invite.
Furthermore, the weights for each tracer are methodology dependent, such that parallel analyses of the same dataset may vary in their expectations for cosmic variance.
For example, our analysis of the SALT sample found the isotropic variance $\sqrt{\text{Tr}(\bm{\Sigma}_{\text{cos}})}$ differed by almost a factor of two (172.1 km s$^{-1}$ to $310.0$ km s$^{-1}$) between the Peery MVE and the Nusser MLE.

A direct comparison between studies would require deploying the exact methods on both datasets to isolate systematic discrepancies and identify points of divergence.
It is not currently standard to provide code that fully reproduces an analysis, but such a convention would enable more accurate and meaningful meta-analyses in the future.
To that end, we publish all code used in this work at \url{https://github.com/ado8/hsf_bulkflows}.
For the contemporary reader insisting upon an enumeration of studies finding bulk flows consistent with or anomalous under $\Lambda$CDM, we recommend Sections 9.2 and 9.3 of \citet{tsagas26}.

\subsection{Future Work}
\label{sec:future_work}

As stated in Section \ref{sec:validation}, the most realistic mock catalogues use N-body simulations performed in theoretically motivated cosmologies and account for observational and selection effects.
Our mock catalogues do not incorporate the non-linear effects produced by structure formation, are limited to two fiducial cosmologies, and do not incorporate selection effects.

\subsubsection{Computational Challenges}
The N-body code L-PICOLA \citep{howlett15} has been used in several studies of peculiar velocities and bulk flow \citep[e.g.][]{hellwing18, qin18, qin21, howlett22, whitford23}, but the computational cost of generating mocks would increase by orders of magnitude to become a limiting factor in the validation process.
Further exacerbating this issue, sampling a continuous distribution of $H_0$ and $\Omega_M$ values in mock-generation would significantly expand the parameter space, necessitating a commensurate increase in simulations to maintain statistical power.

\subsubsection{Potential Computational Solution: Caching}
The increase in computational costs could be mitigated by reusing simulations with observers placed at different coordinates or downsampling a list of observable tracers to produce catalogues.
However, any overlap between light-cones or tracers would induce systematic correlations in the results that may invalidate analyses assuming independence.
If the degree of correlation can be parametrized and its effects modelled, the analytic framework may be modifiable to avoid assuming independence, but it is not guaranteed that caching will produce larger \textit{effective} sample sizes with the same amount of computational resources.

\subsubsection{Analytical Challenges}
If a selection function can be defined, applying a simple completeness correction to the weights has been shown to yield results consistent with those found with a volume-complete selection function \citep{li12}.
However, specifying an accurate selection function is not trivial.
Our analysis models selection functions as a prescribed function with a set of sample-level parameters conditioned on distance moduli.
Other options for modelling selection effects are presented in Section \ref{sec:preprocessing}, but typically require working with photometry rather than with distance moduli.
Sophistication could be added to our model, but more sophisticated models already exist.

\subsubsection{Potential Analytical Solution: Hierarchical Bayes}
The BayeSN model \citep{thorp21, mandel22, ward23, grayling24} offers a hierarchical Bayesian framework for forward modelling the spectral energy distribution of a SN~Ia to be conditioned on photometric data.
Bulk flow inference is a hierarchical problem which involves parameters specific to elements within a set (e.g. radial peculiar velocities and distance moduli of tracers) and common parameters shared within subsets (e.g. selection likelihood, tracer spatial distribution, the velocity field).
The Bayesian framework formalises the use of prior information in probabilistic terms to make inferences, particularly important for bulk flows given the sparsity of peculiar velocity catalogues.
BayeSN does not currently support peculiar velocity inference, but implementing that feature would be simpler than encoding photometry-level analyses in our model.

There are several advantages BayeSN offers over traditional light-curve fitting techniques that are relevant in the context of bulk flow inference.
First, covariances between data and parameters of interest are preserved due to their simultaneous inference.
As an example, the covariance matrix for radial peculiar velocities (Equation \ref{eqn:pv_covar}) is a function of distances between tracers.
The traditional method would sequentially calculate distance moduli, then infer peculiar velocities regardless of their likelihood under a derived covariance matrix.
The Bayesian approach would probe the likelihood over a multi-dimensional parameter space, balancing covariant quantities against each other to yield improved estimates of both.
Second, as a hierarchical model, BayeSN supports the inference of shared, population-level parameters like the selection function within a given survey or the scale of non-linear dispersion ($\sigma_\text{NL}$).
Third, as a Bayesian model, prior information like the expected rates of SN~Ia per comoving volume or the density map needed for calculating inhomogeneous Malmquist bias can be incorporated explicitly and transparently.

Extending the current BayeSN model to include peculiar velocities and bulk flows is the most direct path towards a hierarchical Bayesian framework for bulk flow inference.


\section{Acknowledgements}

This work, AD, and KSM are supported by the European Research Council (ERC) under the European Union’s Horizon 2020 research and innovation programme under Grant Agreement No. 101002652 (BayeSN; PI K. Mandel) and Marie Skłodowska-Curie Grant Agreement No. 873089 (ASTROSTAT-II).
BS is supported by National Science Foundation grant AST-1911074.

This publication makes use of data products from the Two Micron All Sky Survey, which is a joint project of the University of Massachusetts and the Infrared Processing and Analysis Center/California Institute of Technology, funded by the National Aeronautics and Space Administration and the National Science Foundation.

The Pan-STARR Surveys (PS1) and the PS1 public science archive have been made possible through contributions by the Institute for Astronomy, the University of Hawaii, the Pan-STARRS Project Office, the Max-Planck Society and its participating institutes, the Max Planck Institute for Astronomy, Heidelberg and the Max Planck Institute for Extraterrestrial Physics, Garching, The Johns Hopkins University, Durham University, the University of Edinburgh, the Queen's University Belfast, the Harvard-Smithsonian Center for Astrophysics, the Las Cumbres Observatory Global Telescope Network Incorporated, the National Central University of Taiwan, the Space Telescope Science Institute, the National Aeronautics and Space Administration under Grant No. NNX08AR22G issued through the Planetary Science Division of the NASA Science Mission Directorate, the National Science Foundation Grant No. AST-1238877, the University of Maryland, Eotvos Lorand University (ELTE), the Los Alamos National Laboratory, and the Gordon and Betty Moore Foundation.

This research has made use of NASA’s Astrophysics Data System.

We acknowledge the usage of the HyperLeda database (\url{http://leda.univ-lyon1.fr}).

This research has made use of the SIMBAD database, operated at CDS, Strasbourg, France

This research has made use of the NASA/IPAC Extragalactic Database (NED), which is funded by the National Aeronautics and Space Administration and operated by the California Institute of Technology.

UKIRT is owned by the University of Hawaii (UH) and operated by the UH Institute for Astronomy.
When (some of) the data reported here were obtained, the operations were enabled through the cooperation of the East Asian Observatory.

The ZTF forced-photometry service was funded under the Heising-Simons Foundation grant
\#12540303 (PI: Graham).

This work has made use of data from the European Space Agency (ESA) mission {\it Gaia} (\url{https://www.cosmos.esa.int/gaia}), processed by the {\it Gaia} Data Processing and Analysis Consortium (DPAC, \url{https://www.cosmos.esa.int/web/gaia/dpac/consortium}).
Funding for the DPAC has been provided by national institutions, in particular the institutions participating in the {\it Gaia} Multilateral Agreement.

We acknowledge ESA Gaia, DPAC and the Photometric Science Alerts Team (\url{http://gsaweb.ast.cam.ac.uk/alerts}).

Funding for the Sloan Digital Sky Survey (SDSS) has been provided by the Alfred P. Sloan Foundation, the Participating Institutions, the National Aeronautics and Space Administration, the National Science Foundation, the U.S. Department of Energy, the Japanese Monbukagakusho, and the Max Planck Society. The SDSS Web site is \url{http://www.sdss.org/}.

The SDSS is managed by the Astrophysical Research Consortium (ARC) for the Participating Institutions. The Participating Institutions are The University of Chicago, Fermilab, the Institute for Advanced Study, the Japan Participation Group, The Johns Hopkins University, the Korean Scientist Group, Los Alamos National Laboratory, the Max-Planck-Institute for Astronomy (MPIA), the Max-Planck-Institute for Astrophysics (MPA), New Mexico State University, University of Pittsburgh, University of Portsmouth, Princeton University, the United States Naval Observatory, and the University of Washington.

This work is based on observations made by UKIRT, ATLAS, UH2.2, Subaru, and ASAS-SN.
The authors wish to recognise and acknowledge the very significant cultural role and reverence that the summits of Haleakalā, Mauna Loa, and Maunakea has always had within the indigenous Hawaiian Community.
We are most fortunate to have the opportunity to conduct observations from these mountains.
\section{Artificial Intelligence Disclosure Statement}
AD provides the following AID statement \citep{weaver24}.

\textit{Artificial Intelligence Tool}: Gemini v3;
\textit{Information Collection}: Gemini identified the citation for Equation \ref{eqn:erfc_integral};
\textit{Methodology}: Gemini indicated that a factor of $\Om$ should be included in Equation \ref{eqn:rad_pv_pointsource} and
suggested Gauss-Hermite quadrature approximation in Appendix \ref{appendix:MVE_mod};
\textit{Data Analysis}:
Gemini verified Equations \ref{eqn:weight_bias} and \ref{eqn:u_exact_bias},
verified the derivation in Appendix \ref{appendix:selection_norm},
identified the analytic expression for the radial component of the volume integral in Appendix \ref{appendix:bulk_flow_integral},
suggested the reformulation in Equation \ref{eqn:no_num_inst} to prevent numerical instabilities,
refactored AD's code for both readability and performance,
and refactored the weight calculations in the Whitford code into \texttt{JAX};
\textit{Visualisation}: Gemini adjusted the plotting code used to generate Figures \ref{fig:mocks_symm_uniform_planck_planck} to \ref{fig:mocks_declim_pointsource_planck_SH0ES_symm} and Figure \ref{fig:watkins_estimator} to avoid overlapping labels and improve tick marker readability.

\bibliographystyle{mnras.bst}
\bibliography{bibliography.bib}

\appendix
\section{Systematic Biases in Distance Inference}
\label{appendix:univ_dist_biases}
A number of factors make it difficult to measure the luminosity distances needed for accurate inference.
The following effects apply to all distance indicators that provide an estimate ($d_\text{est}$) of the true luminosity distance ($d_\text{true}$), often in the form of an observed distance modulus ($\mu_\text{obs}$) with Gaussian uncertainty $\sigma_\mu$.
Defining $\kappa \equiv \frac{\ln{10}}{5}$, and a residual term $\epsilon \sim \mathcal{N}(0,~\sigma_\mu^2)$, the relationship between the observed, estimated, and true quantities are as follows:
\begin{equation}
    \begin{split}
        \mu_\text{obs} =& 5\log_{10}(d_\text{true}/1~\text{Mpc})+25 + \epsilon \\
        d_\text{est} =& 10^{\mu_\text{obs}/5-5}~\text{Mpc} = d_\text{true}\exp(\kappa\epsilon)
    \end{split}
\end{equation}
Quantifying the biases assumes $\sigma_\mu$ applies to all targets in a sample, which may ignore differences due to data acquisition or reduction.
As an example, distance estimates based on multiple measurements (e.g., galaxies hosting multiple SNe, clusters with distance estimates from each galaxy) will provide systematically more precise distance estimates.

\subsection{Log-Normal Bias}
\label{appendix:log-normal}
Averaging over all realisations of $\epsilon$, $\langle d_{\text{est}} \rangle = d_\text{true}\langle \exp(\kappa \epsilon) \rangle \neq d_\text{true}$.
\citet{watkins23} derive the ratio between $d_\text{true}$ and the average $d_\text{est}$, which is inverted if inferring the average $d_\text{true}$ from a single $d_\text{est}$ as in \citet{lynden-bell88}
\begin{equation}
    \begin{split}
    \frac{d_\text{true}}{\langle d_\text{est}\rangle}\Bigg|_\text{log-norm}  &=  \exp{\left(-\frac{1}{2}\kappa^2 \sigma_\mu^2\right)} \\
    \frac{\langle d_\text{true} \rangle}{d_\text{est}}\Bigg|_\text{log-norm} &= \exp{\left(\frac{1}{2}\kappa^2 \sigma_\mu^2\right)}.
    \end{split}
\end{equation}
 
This effect has been addressed by using bias-corrected distance or peculiar velocity estimators \citep[e.g.,][]{sorce15, watkins15, hoffman21, qin21gaussian, watkins23} or by working directly with distance moduli or log-distances directly while Monte Carlo sampling \citep[e.g.,][]{nusser99, lavaux16, qin18, valade22}.

\subsection{Homogeneous Malmquist Bias}
\label{appendix:malmquist}
Malmquist bias often refers to a selection effect based on a magnitude, flux, or signal-to-noise ratio (SNR) limit \citep{malmquist22}, but in an extragalactic context \citet{lynden-bell88} used the term to describe a slightly different bias related to differential volume elements scaling with distance.
If galaxies have symmetric odds of randomly scattering near or far in a given distance measurement, a sample of galaxies at a given distance will consist of more galaxies that scattered into the sample from farther away than from near.
The bias is a function of the galaxy number density $n(d_\text{true})$, but can be separated into a homogeneous component that assumes uniform number density, and an inhomogeneous component based on the actual distribution.
The effect of homogeneous Malmquist bias (HMB) is given as
\begin{equation}
        \frac{\langle d_\text{true} \rangle}{d_\text{est}}\Bigg|_\text{HMB}= \exp{\left(3\kappa^2\sigma_\mu^2\right)}.
\end{equation}
which does not include the log-normal bias.

HMB can be addressed through corrective factors or by incorporating geometrically-informed priors on distances as recommended by \citet{desmond25} (e.g., $\pi(r_i) \propto r_i^2$) and implemented in this work.
With a constant comoving density of \sneia{}, the prior can be converted to distance-modulus space as $\pi(\mu) = |\frac{\mathrm{d}r}{\mathrm{d}\mu}|\pi(r) \propto r^3$ if $r = 10^{\mu/5 -5}~\text{Mpc}$ as defined in \citet{desmond25}.
However, when luminosity and comoving distances are not assumed to be equal the Jacobian is no longer strictly linear in $r$ and the prior in distance-modulus space is 
\begin{equation}
    \pi(\mu) \propto r^3 \left(1 + \frac{r H(\zcos)}{c(1+\zcos)}\right)^{-1}
\end{equation}
where $H(z)$ is the Hubble parameter at redshift $z$ and is defined as $H_0\sqrt{\Omega_M(1+z)^3 + (1-\Omega_M)}$ in a flat universe.
The $r^3$ proportionality is recovered when $r$ is much less than the Hubble distance at $\zcos$, with the two approaching equality at $\zcos \approx 0.7$.

\subsection{Inhomogeneous Malmquist Bias}
\label{appendix:IMB}
While HMB quantifies the effects of a spatially uniform distribution of tracers, inhomogeneous Malmquist bias (IMB) is a function of the actual distribution.
With the logarithmic derivative of tracer overdensity with respect to distance $\gamma \equiv \mathrm{d}\ln{n(d_\text{est})}/\mathrm{d}\ln{d_\text{est}}$, a first-order expansion gives the effect of IMB as \citep{lynden-bell88, strauss95}
\begin{equation}
    \frac{\langle d_\text{true} \rangle}{d_\text{est}}\Bigg|_\text{IMB} \approx \exp{(\gamma\kappa^2\sigma_\mu^2)}.
    \label{eqn:IMB}
\end{equation}

Typical peculiar velocity surveys do not measure $\gamma$, but one could iterate towards a solution by using peculiar velocities to infer the distribution of mass, combine that with a galaxy bias factor to estimate $\gamma$, and produce a slightly more bias-corrected set of peculiar velocities.
\citet{boruah22} took a Bayesian approach by using the 2M++ density field \citet{carrick15} as an informative prior.
We do not account for IMB in our data.
\section{Systematic Biases in Bulk Flow Inference}
\label{appendix:MVE_mod}

Equations \ref{eqn:bulk} and \ref{eqn:bulk_alt} define bulk flow as a volumetric integral, which introduces a qualitative mismatch between the quantity to be inferred and the discrete data used to make the inference.
A scalar potential defines a velocity field at each point in space, but the field itself cannot be directly measured.
Instead, we rely on observations of tracers (galaxies or clusters).
Equation \ref{eqn:bulk_flow_sum} describes how the projected bulk flow can be estimated through a discrete weighted average of radial velocity measurements.
The analytic prescription for the various weight terms are detailed in their corresponding references \citep{kaiser88, nusser14, watkins09, peery18}.
In this section we describe systematic issues related to the inferred weights and the peculiar velocity estimates, and the modifications to the Whitford code used to counteract those issues.

\subsection{Weighting}

The Nusser MLE, MVE, and Peery MVE methods have distinct formulations for calculating weights, but all include a common proportionality $w\propto r^{-2}$.
While the comoving distance can be estimated from an observed distance modulus (and accounting for the biases in Appendix \ref{appendix:univ_dist_biases}), the significant uncertainties have led to the convention of using $\zobs$ to calculate $r$.
The uncertainties in $c\zobs$ are orders of magnitude lower than the uncertainties in $r(\mu_{obs})$, which leads to corresponding improvements in statistical precision at the expense of a systematic bias where positive (negative) peculiar velocities lead to greater (lesser) inferred distances and lower (higher) weights.
The magnitude of this bias can be estimated to first order ($\zobs \approx \zcos + u/c$ and $r(\zcos) \approx d_\text{L}(\zcos)\approx c\zcos/H_0$) as
\begin{equation}
    w(\zobs) \approx w(\zcos)\left(1-\frac{2u}{c\zcos}\right).
    \label{eqn:weight_bias}
\end{equation}

In the Whitford code, this only affects the Nusser MLE, since the MVE methods incorporate the $r^{-2}$ weight with an idealized survey with a uniform distribution in $r$ and no correlation between weight and $r$, which is statistically comparable to a uniform distribution in comoving volume with a weighting factor of $r^{-2}$.

We avoid this issue when calculating the Nusser MLE weights by using $\mu_\text{obs}$ to infer $r$ rather than $\zobs$.
This means the systematic biases in Appendix \ref{appendix:univ_dist_biases} must be reckoned.
Furthermore, $\langle r_\text{true}^k \rangle /r_\text{est}^k$ is not just the $k$th power of $\langle r_\text{true} \rangle /r_\text{est}$.
The log-normal bias scales with $k^2$ while the homogeneous and inhomogeneous Malmquist components scale with $k$.
Thus, $\langle r_\text{true}^{-2} \rangle /r_\text{est}^{-2} \approx \exp\left[(-4-2\gamma)\kappa^2\sigma_\mu^2\right]$.
We calculate $r_\text{est}^{-2} = \langle r_\text{true}^{-2}\rangle$ using an assumed cosmology and a corrected distance modulus of $\mu_\text{obs} - 4\kappa\sigma^2_\mu$.

The effect of calculating weights based on distance moduli corrected to be unbiased in $r^{-2}$ or $r$ ($\Delta \mu_\text{corr} = 7.5\kappa\sigma_\mu^2$) is marginal.
The inferred bulk flow moments differ by an average of $\sim 10^{-2}~\text{km s}^{-1}$, with a standard deviation of $\sim 0.5~\text{to}~1~\text{km s}^{-1}$.
There was no trend between the differences and the simulated bulk flows, but it is possible the differences could be correlated with simulation parameters not varied: radial distribution of tracers, survey depth, $\sigma_\mu(\zcos)$, etc.
The uncertainties on the moments are lower when using our nominal correction by an average of $\sim 10^{-1}~\text{km s}^{-1}$, and the scale of the difference is correlated with  uncertainties.

\subsection{Peculiar Velocities}
When inferring a radial peculiar velocity as a function of $\mu_\text{obs}$ and $\zobs$ using Equation \ref{eqn: pv_rad}, untreated biases in distance inference propagate to the peculiar velocity.
To first order, the average estimate has both a multiplicative and additive bias:
\begin{equation}
    \langle u_\text{est} \rangle \approx u_\text{true}\left( 1 - \frac{\zcos}{1+\zcos} \frac{\kappa^2\sigma_\mu^2}{2}  \right) - \frac{c\zcos}{1+\zcos}\frac{\kappa^2\sigma_\mu^2}{2}.
    \label{eqn:u_exact_bias}
\end{equation}
The contributions from the homogeneous and inhomogeneous components of Malmquist bias are not relevant here because the estimate is a function of true values.

The inverse-problem would yield $\langle u_\text{true} \rangle$ given a naively inferred $u_\text{est}$, but it is simpler to either use an alternative estimator, or to treat the bias in distance inference first and use an unbiased estimate of $\zcos$ to infer $u_\text{true}$ with Equation \ref{eqn: pv_rad}.

\subsubsection{The \citet{watkins15} Estimator}
\begin{figure*}
    \includegraphics[width=\textwidth]{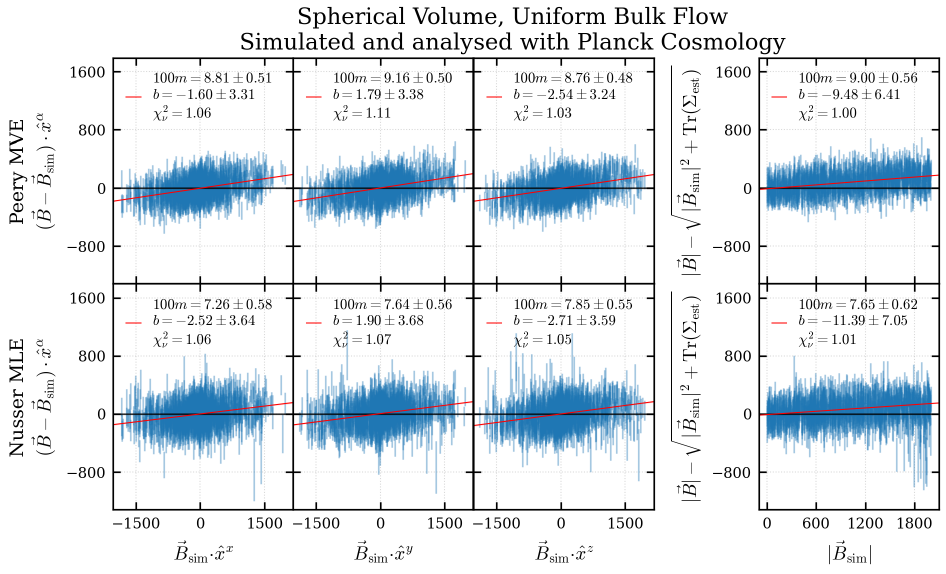}
    \caption{The same format as Figure \ref{fig:mocks_symm_uniform_planck_planck}, but the radial peculiar velocities were estimated through the \citet{watkins15} estimator reproduced in  Equation \ref{eqn:watkins_estimator}. This led to recovered speeds $\sim9\%$ ($\sim7\%$) larger than the simulated values for the Peery MVE (Nusser MLE) method. Similar overestimates are seen when varying the shape of the survey volume, type of bulk flow simulated, and the cosmology assumed for simulation and analysis, although mismatches between the last two terms can increase or decrease the scale of the overestimation.
    }
    \label{fig:watkins_estimator}
\end{figure*}

The Whitford code uses the peculiar velocity estimator and uncertainties from \citet{watkins15},
\begin{equation}
    \begin{split}
    u_\text{est} =& \frac{cz_\text{mod}}{1+z_\text{mod}} \eta \\
    \sigma_u = & \frac{cz_\text{mod}}{1+z_\text{mod}} \frac{\ln(10)}{5} \sigma_\mu
    \end{split}
    \label{eqn:watkins_estimator}
\end{equation}
where $\eta$ is the log-distance ratio provided by the user, described in \citet{watkins15} as $\eta = \ln\left(\frac{cz_\text{mod}}{H_0d_{L}}\right)$, and $z_\text{mod}$ is defined in Equation 2 of the same paper as
\begin{equation}
    z_\text{mod} = \zobs\left[1+\frac{1-q_0}{2}\zobs - \frac{1-q_0-3q_0^2 + 1}{6}\zobs^2\right],
\end{equation}
such that in the absence of peculiar velocities $H_0d_L \approx cz_\text{mod}$ holds to third-order rather than first.
The estimator is derived from $1+z_\text{mod} = (1+H_0d_\text{L})(1+u/c)$, which does not strictly follow from the standard $1+\zobs = (1+\zcos)(1+u/c)$, but allows for the use of $H_0d_\text{L}$ rather than a $\zcos$ inferred from $d_\text{L}$.
This estimator is statistically unbiased ($\langle u_\text{est} - u_\text{true}\rangle \approx 0$) and has Gaussian errors.
However, initial testing following the methods described in Section \ref{sec:validation} demonstrated that the use of this estimator produced overestimated bulk flow speeds as shown in Figure \ref{fig:watkins_estimator}.

\begin{figure}
    \includegraphics[width=0.49\textwidth]{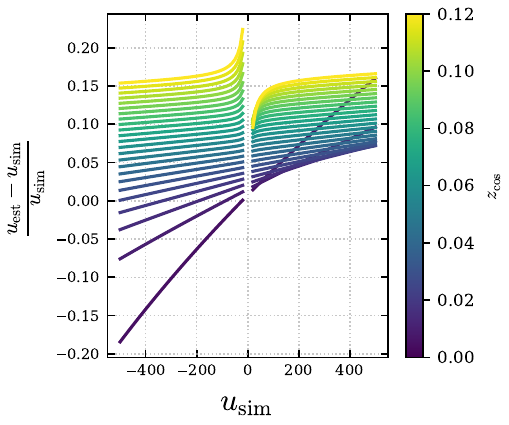}
    \caption{The relative difference between the radial velocity calculated with Equation \ref{eqn:watkins_estimator} ($u_{\text{est}}$) and the simulated value ($u_{\text{sim}}$) is a function of both $u_{\text{sim}}$ and the cosmological redshift ($\zcos$).
    The $u_\text{est}$ values in this plot were calculated using simulated observations with no noise.
    Simulated speeds less than or equal to 10 km s$^{-1}$ are not plotted to avoid the vertical asymptote.
    }
    \label{fig:watkins_bias}
\end{figure}

While the estimator is unbiased with respect to simulated radial velocities, this is only because positive $u_\text{true}$ values are overestimated by about as much as the negative values are underestimated.
The average velocity is unbiased but the average speed is overestimated, as shown in Figure \ref{fig:watkins_bias}.
The magnitude of this bias is roughly proportional to $\zcos$, but there is no redshift at which $u_{\text{est}} \approx u_{\text{true}}$ for all $u_{\text{true}}$.
The spatially uniform distribution of our simulations combined with the $r^{-2}$ weighting in the Peery MVE and Nusser MLE methods places the average weight in our mock catalogues at half the maximum radius, where $(u_{\text{est}} - u_{\text{sim}})/u_{\text{sim}} \approx 0.085$.
This crude estimate is consistent with the slopes seen in Figure \ref{fig:watkins_estimator}.
A more sophisticated but computationally expensive examination could compare the inferred bias over a range of simulated survey depths, with a positive correlation expected if the speed bias of the \citet{watkins15} estimator is the source of the speed bias in the bulk flow.

\subsubsection{Bias Corrected Distance Moduli}
Treating biases at the distance modulus level and using Equation \ref{eqn: pv_rad} to infer radial peculiar velocities will produce estimates that are unbiased to first-order in both velocity and speed, but distributed with non-Gaussian errors.
This is a qualitative departure from, rather than a strict improvement over, the \citet{watkins15} estimator which is unbiased in velocity, biased in speed, and produces Gaussian errors.
Gaussianity is relevant because the MVE and Peery MVE methods include an assumption that the measured peculiar velocity can be modelled as $u_\text{est} \sim \mathcal{N}(u_\text{true},~\sigma_{u,n}^2 + \sigma_\text{NL}^2)$, where $\sigma_{u,n}$ is the measurement uncertainty of the radial peculiar velocity for the $n$th tracer and $\sigma_\text{NL}$ is the uncertainty due to non-linear effects \citep{watkins09}.
However, the Gauss-Markov theorem \citep[reviewed in][]{hallin14} supports the use of generalised least-squares analyses like the MVE and Peery MVE methods in linear models even with non-Gaussian errors as long as the variances and expectation values are specified accurately.
The departure from Gaussian errors means that there may exist a \textit{non-linear} unbiased estimator with lower sampling variance than the MVE and Peery MVE, but does not compromise the mathematical validity of those methods.
Thus, we choose to apply corrective factors to the observed distance moduli when using them to infer distances or peculiar velocities.

Given observables $\mu_\text{obs}$, $\zobs$ and Gaussian uncertainty $\sigma_\mu$, and assumptions for cosmological parameters $H_0$ and $\Omega_M$, our velocity estimator is
\begin{equation}
    \begin{split}
        u_\text{est} =&~ c\left(\frac{1+\zobs}{1+\zcos(\mu_\text{corr},~H_0~,\Omega_M)} -1\right)\\
        \mu_\text{corr} =&~ \mu_\text{obs} + \alpha \frac{\ln(10)}{5}\sigma_\mu^2
    \end{split}
    \label{eqn:new_estimator}
\end{equation}
where the bias-correction factor $\alpha = 3.5$ for our simulations, but only $0.5$ for our pre-processed data, which have already been treated for HMB.
In both cases, the $\gamma$ term for IMB is treated as 0 and $\zcos$ is inferred with \texttt{wcosmo}.

The statistical properties of this estimator are less tractable than those of the \citet{watkins15} estimator.
If variance due to $\zobs$ is neglected the roughly log-normal distribution of $\zcos$ produces a \textit{logit}-normal distribution in $u_\text{est}$ for which the mean and variance have no analytical solution.
That said, the Gaussian distribution in $\mu_\text{obs}$ permits the use of Gauss-Hermite quadrature for approximating the moments of any function of $\mu_\text{obs}$ via \citep[\href{https://dlmf.nist.gov/3.5}{Equations 3.5.15 and 3.5.28};][]{olver10}
\begin{equation}
    \int_{-\infty}^\infty f(x)e^{-x^2}dx = \sum_{k=1}^q w_kf(x_k) + E_q(f).
\end{equation}
where $x_k$ are the nodes of the $q$th order monic Hermite polynomial and  $w_k$ are the corresponding weights \citep[both taken from \href{https://dlmf.nist.gov/3.5}{Table 3.5.10-13} of][]{olver10}, both insensitive to function $f$.
The remainder function $E_q$ does depend on $f$, but shrinks towards 0 as $q$ increases (we use $q=20$).
We approximate the first and second moments of the probability distribution of $u_\text{est}$ from Equation \ref{eqn:new_estimator} to calculate variance, then add a variance due to uncertainty in $\zobs$ to get total variance.

\section{Derivation of the Distance-Prior Normalisation}
\label{appendix:selection_norm}

The prior given in Equation \ref{eqn:r_prior} is reproduced here for convenience:
\begin{equation*}
    \pi(r) = Cr^2 \text{erfc}\left(\frac{r - R_{\text{cut}}}{\sigma_{\text{cut}}}\right).
\end{equation*}
The normalisation constant $C$ enforces $\int_0^\infty \pi(r) \mathrm{d}r = 1$ and uses the following indefinite integral, derived through integration by parts, the derivative $\mathrm{d}(\text{erfc}(x))/\mathrm{d}x = -2\exp(-x^2)/\sqrt{\pi}$ \citep[trivially derived from \href{https://dlmf.nist.gov/7.2}{Equation 7.2.2} of][]{olver10}, and the definition of the upper incomplete Gamma function \citep[\href{https://dlmf.nist.gov/8.2}{Equation 8.2.2};][]{olver10}.
\begin{equation}
    \begin{split}
        I(p) =& \int \text{erfc}(x) x^{p} \mathrm{d}x \\
        =& \frac{x^{p+1}}{p+1} \text{erfc}(x) + \frac{1}{(p+1)\sqrt{\pi}} \Gamma \left(\frac{p}{2}+1,~x^2 \right)
    \end{split}
    \label{eqn:erfc_integral}
\end{equation}
where $p>-2$, $p\neq -1$, and the constant of integration is omitted since it does not affect the definite integral.
Using $\sigma = \sigma_{\text{cut}}$ and $R = R_\text{cut}$ for brevity and substituting $x = (r - R)/\sigma$, the unnormalised integral becomes
\begin{equation}
    \int(x\sigma + R)^2 \text{erfc}(x) \sigma\mathrm{d}x = \sigma[\sigma^2I(2) + 2\sigma RI(1) +R^2 I(0)]
\end{equation}
We expand the right-hand side and combine like terms such that the indefinite integral can be expressed as $C_\text{erfc}(x) + C_\Gamma(x)$ where 
\begin{equation}
    C_\text{erfc}(x) = \frac{\sigma}{3}(\sigma^2 x^3 + 3\sigma R x^2 + 3R^2 x) \text{erfc}(x)
\end{equation}
and
\begin{equation*}
        C_\Gamma(x) = -\frac{\sigma}{3\sqrt{\pi}}[\sigma^2\Gamma(2,~x^2) + 3\sigma R \Gamma(3/2,~x^2) + 3R^2 \Gamma(1,~x^2)].
\end{equation*}
Using the recurrence relation $\Gamma(a+1,~z) = a\Gamma(a,~z) + z^ae^{-z}$ \citep[\href{https://dlmf.nist.gov/8.8}{Equation 8.8.2};][]{olver10}, special values $\Gamma(0,~z^2) = \sqrt{\pi}\text{erfc}(z)$ and $\Gamma(1,~z) = e^{-z}$ \citep[\href{https://dlmf.nist.gov/8.8}{Equations 8.8.4-5};][]{olver10}, an erfc term appears in $C_\Gamma(x)$ which we move to $C_{\text{erfc}}(x)$.
The two become
\begin{equation}
    \begin{split}
        C_\text{erfc}(x) =& \left(\frac{1}{3}\sigma^3 x^3 + \sigma^2 R x^2 + R^2 x + \frac{1}{2}\sigma^2R \right) \text{erfc}(x) \\
        C_\Gamma(x) =& \frac{\sigma e^{-x^2}}{3\sqrt{\pi}}(\sigma^2 x^2 + 3\sigma R x + \sigma^2 + 3R^2 ).
    \end{split}
    \label{eqn:r_prior_indefinite}
\end{equation}
These equations can be used to calculate the normalisation constant $C = (C_\text{erfc} + C_\Gamma)^{-1}$ for arbitrary limits of integration if, for example, one wanted to implement a selection function with a minimum and/or maximum $r$ value.
Evaluating the definite integral from $r = 0 \rightarrow x=-R/\sigma$ to $r=x=\infty$,
\begin{equation}
    \begin{split}
    C_\text{erfc}(x)|_{-R/\sigma}^\infty =&
        \left(\frac{3\sigma^2 R + 2R^3}{6}\right)
        \text{erfc}\left(-\frac{R}{\sigma}\right) \\
    C_\Gamma(x)|_{-R/\sigma}^\infty =&
        \frac{\sigma\left(\sigma^2+R^2\right)}{3\sqrt{\pi}}
        \exp\left[-\left(\frac{R}{\sigma}\right)^2\right].
    \end{split}
\end{equation}

The integer power-law index in Equation \ref{eqn:r_prior} enables the use of Equation \ref{eqn:erfc_integral} and special values for the upper incomplete Gamma function.
We did not conduct a comprehensive literature search to see if an exact and finite expression for $C$ exists for non-integer power-law indices.
\section{Analytic Bulk Flows from Point-Source Masses}
\label{appendix:bulk_flow_integral}
The integral in Equation \ref{eqn:tidal_bulk_flow} describes the bulk-flow expected from a point-source mass, and is reproduced here for convenience:
\begin{equation*}
    \begin{split}
        \vec{B} \cdot \hat{x}^{\alpha} =& -\frac{2 G f m_\text{sim}}{H_0\Omega_M V}\int_V \mathrm{d}V \frac{(\vec{r} - \vec{r}_{\text{sim}}) \cdot \hat{r}}{|\vec{r} - \vec{r}_{\text{sim}}|^3} \hat{r} \cdot \hat{x}^\alpha \\
        =&  -\frac{2 G f m_\text{sim}}{H_0 \Omega_M V} I_V(\vec{r}).
    \end{split}
\end{equation*}
This integration can be performed in any coordinate system, but the integrand features symmetries that can be exploited.

Spherical coordinates are a natural choice given measurements in right ascension ($a$), declination ($\delta$), and comoving distance ($r$) within a maximum distance of $R$.
Defining scalar projection $p = \vec{r}_{\text{sim}} \cdot \hat{r}$ (which is notably invariant with $r \equiv|\vec{r}|$), the difference vector $\vec{D} \equiv \vec{r} - \vec{r}_{\text{sim}}$ has a magnitude of $D \equiv |\vec{D}| = \sqrt{r^2 + \rsim^2 - 2rp}$ which has a partial derivative with respect to $r$ of $\frac{\partial D}{\partial r} = \frac{r-p}{D}$.
With the aim of using integration by parts and analytically reformulating the radial integral, we use the relation
\begin{equation}
    \frac{(\vec{r} - \vec{r}_{\text{sim}}) \cdot \hat{r}}{|\vec{r} - \vec{r}_{\text{sim}}|^3} = \frac{r-p}{D^3} = -\frac{\partial}{\partial r} \left( \frac{1}{D} \right)
\end{equation}
which can be substituted into the integral.
With a volume element of $\mathrm{d}V = r^2\cos{\delta}\mathrm{d}r\mathrm{d}\delta \mathrm{d}a$, the volume integral in spherical coordinates can be expressed as a surface integral over the unit sphere ($\Omega$) and a radial integral,
\begin{equation}
    I_V(\vec{r})
    = \int_\Omega \left[\int_0^{R} - \frac{\partial}{\partial r}\left(\frac{1}{D}\right) r^2 dr\right] \hat{r}\cdot \hat{x}^\alpha\cos{\delta} \mathrm{d}a \mathrm{d}\delta.
\end{equation}
The radial integral in the brackets ($I_{\text{rad}}(p)$) can be treated with integration by parts
\begin{equation}
    I_{\text{rad}}(p) = -\frac{r^2}{D} \Bigg|_0^R + \int_0^R
  \frac{2r}{D} \mathrm{d}r.
\end{equation}
Given $\mathrm{d}D = \mathrm{d}r (r-p)/D$, the remaining integral can be rewritten as two simpler integrals by adding and subtracting $2p$ in the numerator,
\begin{equation}
  \int \frac{2r}{D}\mathrm{d}r =
      2\int \mathrm{d}D + 2p\int \frac{\mathrm{d}r}{\sqrt{r^2 + \rsim^2 - 2rp}}.
\end{equation}
Completing the square in the root term ($(r-p)^2 + \rsim^2 - p^2$) allows us to use a substitution of $r-p = \sqrt{r^2_\text{sim} - p^2}\tan\theta$ and $\mathrm{d}r = \sqrt{r^2_\text{sim} - p^2} \sec^2\theta \mathrm{d}\theta$ to rewrite the integral as $\int \sec\theta \mathrm{d}\theta = \ln|\sec\theta + \tan \theta|$ \citep[\href{https://dlmf.nist.gov/4.26}{Equations 4.26.5} and \href{https://dlmf.nist.gov/4.23}{4.23.42};][]{olver10}
\begin{equation}
    \int \frac{\mathrm{d}r} {\sqrt{(r-p)^2 + r^2_\text{sim} - p^2}} = \ln |r-p + \sqrt{(r-p)^2+\rsim^2 - p^2}|.
    \label{eqn:inv_root_integral}
\end{equation}
neglecting constant terms degenerate with the elided constant of integration.
The root simplifies to $D$, so the indefinite integral
\begin{equation}
    \int \frac{2r}{D}\mathrm{d}r = 2D + 2p \ln |r - p + D|.
\end{equation}
Evaluating $I_{\text{rad}}(p)$ at the limits of integration,
\begin{equation}
     I_{\text{rad}}(p) =  -\frac{R^2}{D_R} + 2D_R + 2p \ln \left( \frac{D_R + R + \rsim}{D_R - R + \rsim} \right)
\end{equation}
where $D_R \equiv D|_{r=R} = \sqrt{R^2 + \rsim^2 - 2Rp}$.
The fraction in the logarithmic term was reformulated to avoid a numerical instability as
$p \rightarrow \pm \rsim$,
which would cause
$\rsim \mp p \rightarrow 0$
and
$D_R \rightarrow |\rsim \mp R|$.
Since $D_R^2 - (R-p)^2 = \rsim^2 - p^2$, the original fraction
\begin{equation}
    \frac{D_R + (R-p)}{\rsim - p} =  \frac{\rsim + p}{D_R - (R - p)} = \frac{D_R + R + \rsim}{D_R - R + \rsim}.
    \label{eqn:no_num_inst}
\end{equation}
The left and middle parts of the equation follow from the identical differences of squares, but the middle part still encounters numerical instabilities as $p \rightarrow -\rsim$.
The proportionality permits summing the numerators and denominators without altering the value of the fraction as is done in the right part of the equation, which approaches $\frac{2\rsim \mp R+R}{2\rsim \mp R - R}$ as $p \rightarrow \pm \rsim$.
This formulation only suffers from numerical instabilities if $\rsim \rightarrow R$.

The bulk flow projected along the $\alpha$ Cartesian axis is thus
\begin{equation}
    \vec{B} \cdot \hat{x}^\alpha =-\frac{2Gfm_{\text{sim}}}{H_0 \Omega_MV} \int_\Omega I_{\text{rad}}(p) \hat{r}\cdot \hat{x}^{\alpha} \cos{\delta} \mathrm{d}a \mathrm{d}\delta.
\end{equation}

\label{lastpage}
\end{document}